\documentclass[aip,jcp,amsmath,amssymb,reprint]{revtex4-1}
\pdfoutput=1
\usepackage{graphicx}
\usepackage{cases}
\usepackage{upgreek}
\usepackage{fancyhdr}
\newcommand \rmm[1]  { \textrm{#1} }   

\begin{document}
\title{Finite electric displacement simulations of polar ionic solid-electrolyte interfaces: Application to NaCl(111)/aqueous NaCl solution}
\author{Thomas Sayer}
\email{tes36@cam.ac.uk}
\affiliation{Department of Chemistry, University of Cambridge, Cambridge CB2 1EW, United Kingdom}
\author{Michiel Sprik}
\affiliation{Department of Chemistry, University of Cambridge, Cambridge CB2 1EW, United Kingdom}
\author{Chao Zhang}
\affiliation{Department of Chemistry-\AA{}ngstr\"{o}m Laboratory, Uppsala University, L\"{a}gerhyddsv\"{a}gen 1, BOX 538, 75121, Uppsala,
Sweden}

\begin{abstract}
Tasker type III polar terminations of ionic crystals carry a net surface charge as well as a dipole moment and are fundamentally unstable. In contact with electrolytes, such polar surfaces can be stabilized by adsorption of counter ions from solution to form electric double layers (EDLs). In a previous work (J.~Chem.~Phys {\bf 147}, 104702 (2017)) we reported on a classical force field based molecular dynamics study of a prototype model system namely a NaCl(111) slab interfaced with an aqueous NaCl solution on both sides. A serious hurdle in the simulation is that the finite width of the slab admits an electric field in the solid perturbing the theoretical charge balance at the interface of semi-infinite systems (half the surface charge density for NaCl(111)). It was demonstrated that the application of a finite macroscopic field $E$ cancelling the internal electric field can recover the correct charge compensation at the interface. In the present work, we expand this method by applying a conjugate electric displacement field $D$. The benefits of using $D$ instead of $E$ as the control variable are two fold: it does not only speed up the convergence of the polarization in the simulation but also leads to a succinct expression for the biasing displacement field involving only structural parameters which are known in advance. This makes it feasible to study the charge compensating phenomenon of this prototype system with density functional theory based molecular dynamics (DFTMD), as shown in this work. 
\end{abstract}

\date{\today}
\maketitle

\section{Introduction} \label{sec:intro}
Termination of ionic crystals can leave the solid with a surface carrying a net charge. While generally less stable when compared to (low index) uncharged surfaces, a net surface charge can still be accommodated, provided the termination does not also create a surface dipole moment. This is the non-polar type II termination in the classification of Tasker\cite{Tasker1979}. The familiar example is the $(111)$ surface of CaF$_2$ (fluorite). In contrast the $(111)$ termination of NaCl (rocksalt) is fundamentally polar and unstable. This is the type III termination in the classification of Tasker\cite{Tasker1979}. Other examples of type III rocksalt surfaces are the $(111)$ surfaces of MgO and NiO. The $(0001)$ surfaces of the corundum (Al$_2$O$_3$, Fe$_2$O$_3$) and wurtzite (ZnO) structure are also type III polar surfaces. For an extensive review of polar surfaces, we refer the reader to the 2008 review by Goniakowski et al.\cite{Goniakowski2008}.  

Type III polar surfaces can be stabilized by a surface reconstruction which eliminates the dipole moment. The reconstruction is necessarily non-stoichiometric as it must change the net surface charge\cite{Goniakowski2008}.  For polar interfaces with vacuum, the non-stoichiometric construction is often observed to occur by removal or addition of ions. Polar solid-solid interfaces also undergo an electronic reconstruction. For polar surfaces in contact with an electrolyte, the compensating charge can be provided by an exchange of ions with the electrolyte. In a previous publication we have investigated such an ionic solid/electrolyte interface, NaCl(111)/NaCl(aq), using classical force field based molecular dynamics (FFMD) simulation\cite{Sayer2017}. This paper will be referred to as paper I. In the present contribution we return to the model system of paper I now also applying density functional theory based molecular dynamics (DFTMD). 

Atomistic modelling of type III interfaces is a major challenge. The key reason is the slab geometry used in modelling. Slabs of finite width can sustain the electric field created by the polarization and frustrate the reconstruction observed in semi-infinite systems. For physical nanosystems (thin films, nanoparticles) this behaviour is real and of great technical as well as practical interest\cite{Noguera2013}. In simulation studies aiming to understand the fate of polar surfaces of semi-infinite crystals, this effect must however be regarded as a finite size error. For typical model system dimensions used in atomistic simulation this finite size effect can be rather serious. 

A finite width slab would be a better representation of a semi-infinite system if it were possible to cancel out the internal electric field. This suggests application of an appropriate bias perpendicular to the surface. Finite external electric fields are relatively easy to implement for slabs in vacuum under open boundary conditions. The challenge is to apply a field to a polar crystalline slab immersed in an electrolyte maintaining full periodic boundary conditions in parallel as well normal directions and without inserting false vacuum spacers. This was achieved in paper I using the finite electric field methods developed by Vanderbilt and colleagues for treating periodic supercells of ferroelectric solids and multilayer systems\cite{Stengel2009a,Stengel2009b,Sprik2018}.        

This compensating field method was already introduced in Ref.~\citenum{Zhang2016b}. There, it was used to study the electrostatics of the compact electric double layer (EDL) formed at the interface between a high concentration aqueous electrolyte (NaCl) and a hard wall with a fixed surface charge. The system was made 3D periodic by introducing a second wall of opposite charge separated from the first wall by a vacuum space (a.k.a. the insulator). These walls are a simple model of a polarized insulating slab. Again, because of the finite width, the slab admits an electric field. The result is an uncompensated EDL with a net finite charge as can be shown by a simple application of the Maxwell interface theorem~\cite{Zhang2016b}. We note that this finite size error must be distinguished from the interaction of a polarized slab with its images under periodic boundary conditions. Polarization of a finite width isolated slab (open boundary conditions) also induces internal electric fields leading equally to an EDL with a net charge. 

\begin{figure*}[t]
\begin{center}
\includegraphics[scale=0.3]{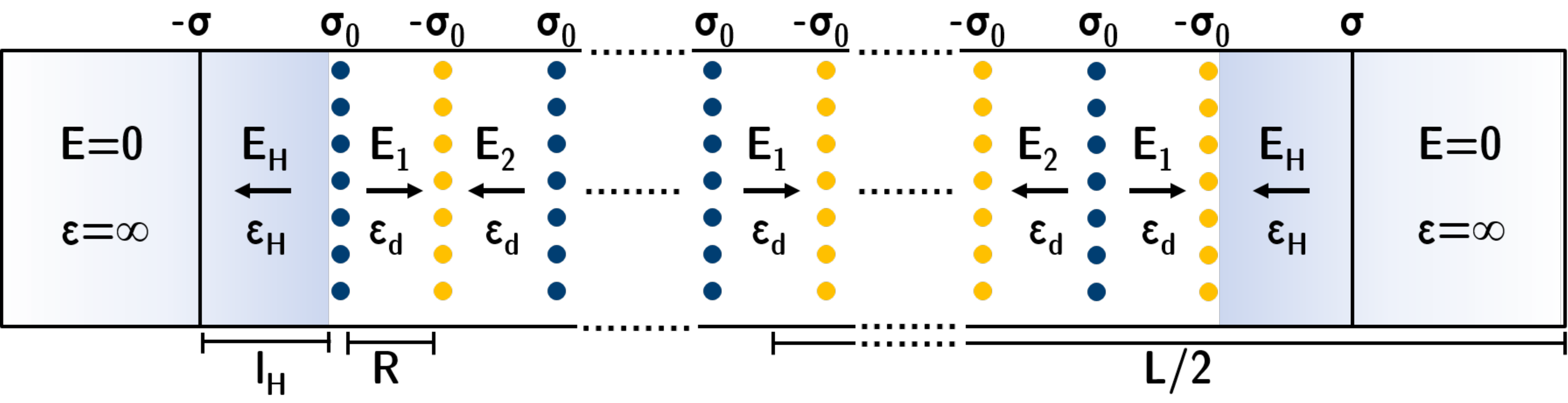}
\end{center}
\caption{Stern model of the ICS from Paper I. The (absolute) surface charge density of a polar surface is $\sigma_0$ and the compensating charge induced in the electrolyte solution is $\sigma$. The solid slab is separated from the electrolyte on both sides by Helmholtz layers. The dielectric constants of the Helmholtz layers and polar solid are $\epsilon_\text{H}$ and $\epsilon_\text{d}$ respectively. The periodic box size is $L$, the width of Helmholtz layer is $l_\text{H}$ and the thickness of a layer in polar solid is $R$. The arrows indicate the convention for the sign of the uniform electric fields in the Helmholtz layers and crystal segments. We additionally note that $E_1$ is in the middle of the slab for $(n+1)/2$ an odd integer, otherwise we have $E_2$. This does not influence the result for the ICS, but will matter for the ECS discussion of Sec.~\ref{sec:aimd}.} \label{fig:stern1}
\end{figure*}

The electrostatics of the model systems in paper I as determined by FFMD were analyzed in detail by comparison to analytic results for the continuum model shown in Fig.~\ref{fig:stern1}. The system is periodic with the solid in the center. The zones of electrolyte on the left and right hand side are part of the same layer of electrolyte intersected by the boundaries of the supercell. The key simplification of the model in Fig.~\ref{fig:stern1} is that the electrolyte is treated as a continuum  with infinite dielectric constant separated from the surface of the solid slab by a layer with a finite dielectric constant $\epsilon_{\mathrm{H}}$. This polarizable continuum represents the compact Stern layer in a high concentration solution (no diffuse layer). $\sigma$ is the ionic surface charge density of the Helmholtz plane. It is the response of the electrolyte screening the surface charge distribution $\sigma_0$ of the ionic solid. In Fig.~\ref{fig:stern1} the atomic point charges of the (111) plane has been smeared out into a homogeneous surface charge distribution $\pm \sigma_0$. The solid consists of $n$ of these alternating charge planes with $n$ as an odd integer. There are three geometric parameters in the model, the width $l_{\textrm{H}}$ of the Stern layer, the spacing $R$ between the charge planes and the repeat length $L$ of the periodic boundary conditions normal to the surface. The width of the solid slab equals $nR$. This leaves a width of $L-2l_{\textrm{H}}-nR$ for the zone of electrolyte. 

With $\sigma_0$ fixed and the potential across the solid slab tuned to zero by an appropriate biasing potential (see method section below) the central variable in the continuum model is the electrolyte surface charge $\sigma$.  The model is a piece wise homogeneous system and $\sigma$ can be determined by application of the Maxwell interface theorem. The solution derived in paper I is  
 \begin{equation}
\label{eq:sigman}
 \sigma_\rmm{CNC} = \frac{n+1}{2n} \sigma_0
\end{equation}
where CNC stands for Compensating Net Charge. The CNC nomenclature was chosen as a generalization of the Zero Net Charge (ZNC) condition restoring the charge balance in the double layers of Ref.~\citenum{Zhang2016b}. Indeed setting $n=1$
in Eq.~\ref{eq:sigman} gives $\sigma_{\rmm{CNC}} = \sigma_0$. For $n \gg 1$ (the thick polar slab limit)  $\sigma_{\rmm{CNC}} = \sigma_0/2$ (Eq.~\ref{eq:sigman} will be derived again in Sec.~\ref{sec:stern}). Eq.~\ref{eq:sigman} interpolates between the neutral double layer system studied in Ref.~\citenum{Zhang2016b} and that of the compensating charge predicted by the Tasker rule for a semi-infinite $(111)$ polar rocksalt surface. This is to be expected. However, less obvious is that the semi-infinite limit is reached gradually. The number of planes in the cross over region is predicted to be independent of any geometric or dielectric parameters. Eq.~\ref{eq:sigman} must be regarded as a residual finite width effect persisting even after the average internal electric field in the slab has been cancelled by an external bias. 

The FFMD results of paper I on the NaCl(111)/NaCl(aq) model system are in excellent agreement with the predictions of Eq.~\ref{eq:sigman}, confirming that it takes a system with a minimum of 15 ionic planes to reduce the compensating electrolyte charge to the half charge Tasker limit. Unfortunately, such system sizes exceed what is feasible in DFTMD. However, in view of the apparently parameter free $n$ dependence of Eq.~\ref{eq:sigman} it seemed acceptable to us to reduce $n$ to a minimum value of $n=3$, which according to Eq.~\ref{eq:sigman} would give a compensation charge of $\sigma= 2 \sigma_0/3$. This is approximately midway between the $\sigma=\sigma_0$ double layer limit and the $\sigma = \sigma_0/2$ semi-infinite limit. We know from paper I that the FFMD system satisfies this relation. The question is whether the DFTMD simulation does as well. Anticipating our results, the DFTMD and FFMD seem to agree upon the value of the net compensating charge while admitting significant differences in the structure of the interface.   

The present contribution goes beyond paper I in a further more important aspect. This concerns the feasibility of applying the constant field method in DFTMD simulation. The constant field that is imposed in paper I is the macroscopic Maxwell field $\bar{E}$. $\bar{E}$ is a control parameter in the constant field Hamiltonian (see Sec.~\ref{sec:hssv}) and what is needed is the particular field $\bar{E}$ cancelling the internal electric field across the solid slab. This field was referred to in paper I as $\bar{E}=E_{\textrm{CNC}}$.  $E_{\textrm{CNC}}$ was empirically determined by searching for the $\bar{E}$ which eliminates the potential drop over the polar crystal slab. This requires a series of finite $\bar{E}$ calculations and so the resulting computational overhead makes it prohibitive for DFTMD simulations.

Dielectric theory defines a second field conjugate to the Maxwell field $E$. This is the dielectric displacement $D$ related to the Maxwell field $E$ as
\begin{equation}\label{eq:defD}
D=E+4\pi P
\end{equation}
where $P$ is the polarization (as in previous publications Gaussian electric units are used). The quantities $D,E$ and $P$ in Eq.~\ref{eq:defD} are a short hand notation for the component of the corresponding vector fields perpendicular to the slab (see Fig.~\ref{fig:stern1}). As an alternative to the constant $E$ method Stengel, Spaldin and Vanderbilt (SSV) also developed a constant $D$ method~\cite{Stengel2009a}. As the SSV constant $E$ and $D$ methods were intended for application in DFT based electronic structure calculation, the methods were formulation in terms of extended Hamiltonians coupling the polarization to a uniform Maxwell field or dielectric displacement~\cite{Stengel2009a,Stengel2009b}. The crucial step in this development was the representation of electronic polarization in terms of a Berry phase, as in the modern theory of polarization~\cite{King-Smith:1993prb,Resta:1994rmp,Resta:2007ch}. The same DFT approach will be applied in the present work. However, the extension term can also be used in combination with an FFMD Ewald Hamiltonian. 

It was shown in in Ref.~\citenum{Zhang2016b} that keeping $D$ fixed is a more efficient method for determining $E_{\textrm{CNC}}$ as will be explained again below. The constant $D$ scheme was not used in paper I and this will be taken up in the present paper. Another proposed advantage is that while $E_\rmm{CNC}$ depended on the capacitance of the (Helmholtz) double layer $C_\rmm{H}$ -- the determination of which is usually an objective of the simulation -- the value of $D_\rmm{CNC}$ involved only structural parameters. In this work, we will derive the expression of $D_\text{CNC}$, validate it with FFMD simulations of NaCl(111)/NaCl Solution and then apply it to DFTMD simulations of exactly the same system. The present contribution is predominantly devoted to establishing the methodology for treating periodic systems with spontaneous polarization and to testing its feasibility in the context of DFTMD simulations (theory and implementation). This is the subject of Secs.~\ref{sec:hssv}-\ref{sec:tech}. The focus of property calculation is therefore on the capacitance and dielectric response (Sec.~\ref{sec:results}). A detailed structural analysis is deferred to follow-up publications.

\section{Constant field molecular dynamics} \label{sec:ssvmd} 

\subsection{Constant field hamiltonians} \label{sec:hssv}

In this section we briefly summarize the constant $E$ and constant $D$ method in order to highlight some important technical issues. We first outline the FFMD implementation as used in paper I and Ref.~\citenum{Zhang2016b}. In the present investigation this method is again used for initialization of the DFT simulation and for comparison to the DFT results. The adaption of SSV constant methods for FFMD simulation of aqueous systems is relatively straightforward\cite{Zhang2016a}. The constant-$\mathbf{E}$ Hamiltonian is written as
\begin{equation}
\label{eq:HvdbE}
H_E  = \sum_i^N  \frac{\mathbf{p}_i^2 }{2 m_i} + 
V_{\mathrm{PBC}}\left(\mathbf{r}^N\right) - \frac{\Omega}{8 \pi} \mathbf{E}^2 
- \Omega \mathbf{E} \cdot \mathbf{P}
\end{equation}
and the constant-$\mathbf{D}$ Hamiltonian as
\begin{equation}
\label{eq:HvdbD}
H_D  = \sum_i^N  \frac{\mathbf{p}_i^2 }{2 m_i} + 
V_{\mathrm{PBC}}\left(\mathbf{r}^N\right) 
+  \frac{\Omega}{8 \pi} \left(\mathbf{D}- 4 \pi \mathbf{P} \right)^2 
\end{equation}
where $\mathbf{p}_i = m_i \mathbf{v}_i$ is the momentum of particle $i$ with position vector $\mathbf{r}_i$ and velocity  $\mathbf{v}_i = \dot{\mathbf{r}}_i$. The mass of particle $i$ is $m_i$.  The potential $V_{\mathrm{PBC}}\left(\mathbf{r}^N\right)$ is the potential energy of the $N$ particle system with the electrostatic  interactions evaluated using standard Ewald summation (no surface term)\cite{Caillol1994,DeLeeuw1980a}. $\Omega$ is the volume of the periodic supercell.

\subsection{Polarization in ionic systems} \label{sec:spcpol}
Polarization in a periodic system is multivalued in nature~\cite{King-Smith:1993prb,Resta:1994rmp,Resta:2007ch}. For example, whenever a unit of charge $e$ wraps across the supercell boundary in the $z$ direction, there is a jump in the polarization. This jump $\Delta_q P$, termed the `quantum of polarization', is a simple function of the supercell dimensions  
\begin{equation}
\label{eq:DqP1}
\Delta_q P = \frac{eL}{\Omega}  
\end{equation}
Eq.~\ref{eq:DqP1} is valid for any orthorhombic supercell of ions and electrons. For our model system of a rigid polar crystal oriented perpendicular to the $z$ axis (Fig.~\ref{fig:stern1}) $\Delta P_q$ can be equated to 
\begin{equation}
\label{eq:DqP0}
\Delta_q P = \frac{\sigma_0}{N_0} 
 \end{equation}  
and $N_0$ is the number of ions in a charged plane. Eq.~\ref{eq:DqP0} is the more convenient expression to be used later in the calculations.

 Polarization in periodic systems is defined modulo the quantum of polarization $\Delta_q P$. This uncertainty is reflected in the constant $\mathbf{E}$ Hamiltonian of Eq.~\ref{eq:HvdbE} (also called the electrostatic enthalpy) which is now also multiply valued. Adopting the language of mathematics the energy surface consists of a stack of branches, each branch defined by a multiple of $\Delta_q P$. In contrast, the constant $\mathbf{D}$ is free from this ambiguity in principle, provided the $\Delta_qP$ fo a given branch of polarization is deducted again from the displacement field. The branches collapse on a single energy surface given by Eq.~\ref{eq:HvdbD}. Indeed the quantity in the brackets of $H_D$ is the electrostatic field (see Eq.~\ref{eq:defD}). $q \mathbf{E}$ is a force on a particle with charge $q$ and therefore uniquely defined. The value of a displacement field in a supercell is therefore also multivalued\cite{Stengel2009b}.     

$\mathbf{E}$ in Eq.~\ref{eq:HvdbE} and $\mathbf{D}$ in Eq.~\ref{eq:HvdbD} are parameters (control variables) and that is why we put bar over them as $\bar{E}$ and $\bar{D}$ in previous publications~\cite{Zhang2016a, Sayer2017}. The polarization $\mathbf{P}$ is a system observable fluctuating in time. For liquid water $\mathbf{P}$ is simply the sum of the molecular dipole moments\cite{Zhang2016a}. For application to ionic solutions all charge, bound (solvent) and mobile (ions) must be included in the polarization\cite{Zhang2016b}. This introduces the confusing complication of multiple values and branches described above. The contribution to polarization of the free ions now depends on the choice of supercell cell\cite{Stengel2009b,Zhang2016b}. During the time evolution, ambiguity is avoided by following an ion when it crosses a supercell boundary. The value of the polarization is therefore fixed by where the ions are in the supercell of choice at the start of the MD trajectory. Formally this definition of polarization can be regarded as an integral of the (total) current and is referred to as itinerant polarization as was already introduced earlier in FFMD simulation of homogeneous ionic solutions\cite{Caillol1989a,Caillol1989b,Caillol1994}. 

The dependence of the polarization on the choice of the unit cell is a necessary artefact. It can be regarded as a gauge. Experimentally observable quantities, such as the electrolyte response charge $\sigma$ should be invariant under a change of supercell. As discussed in Ref.~\citenum{Zhang2016b} the geometry of alternating solid slabs and electrolyte leads to two fundamentally different options for a supercell. The first is a configuration with the slab in the middle and the cell boundaries partitioning the electrolyte in two half's, one to the left and and one to the right of the central slab. This is the configuration depicted in Fig.~\ref{fig:stern1} and was referred in Ref.~\citenum{Zhang2016b} as the insulator centered supercell (ICS). The alternative is placing the electrolyte in the middle and the solid on the side giving an electrolyte centered supercell (ECS).

The crucial difference between the ICS and ECS geometry is that in the ECS the ions are confined by the solid surfaces and cannot escape from the cell. The same supercell can be used to specify the polarization over the entire run eliminating any ambiguity. In the ICS the mobile can cross the cell boundaries and we will have to keep track of where the go in order to refer back to their initial position which was the basis of the definition of polarization. Clearly the MD results obtained in a ICS or ECS must be consistent. This consistency test was the main tool used to validate the FFMD SSV finite field supercell approach in Ref.~\citenum{Zhang2016b}. The freedom of switching between ECS and ICS when this is convenient will be exploited in the present calculation.

\section{Continuum Stern model} \label{sec:stern}

To work out the relevant equations for the continuum approximation of Fig~\ref{fig:stern1} we employ a generalization of the derivation for the more simple EDL model of Ref.~\citenum{Zhang2016b}. We first derive the expression for the full polarization $P$ and electric field $\bar{E}$ for the insulator centered cell (ICS) geometry of Fig.~\ref{fig:stern1}. These equations were already derived in paper I. Then, applying the constitutive relation assumed in the continuum model, we obtain an equation for the compensating internal field $E_{\textrm{CNC}}$ of the polar crystal which will lead to Eq.~\ref{eq:sigman}. Next $E_{\textrm{CNC}}$ is transformed to the corresponding dielectric displacement $D_{\textrm{CNC}}$ according to Eq.~\ref{eq:defD}.   Finally we go over part of the derivation again for the  electrolyte centered cell (ECS) geometry. This setup is more completed compared to the EDL system of Ref.~\citenum{Zhang2016b}. For reasons explained later it will be used as the DFTMD model system. 

\subsection{Insulator centered supercell (ICS)} \label{sec:ics}

\subsubsection{Polarization} \label{sec:pol}
We begin by writing down the net surface charge density $\sigma_{\textrm{H}}$ of the Helmholtz plane separating the electrolyte from the Stern layer. $\sigma_{\textrm{H}}$ is the sum of the charge density induced in the electrolyte and the dielectric material of the Stern layer. The polarization charge density induced in the electrolyte is the $\sigma$ of Eq.~\ref{eq:sigman}. The polarization charge density of the Stern layer is found from  the polarization $P_\rmm{H}$ of the Stern layer by applying the Maxwell interface theorem. Hence we can write
\begin{equation}
\label{eq:maxwell0}
\sigma_\rmm{H} = \sigma - P_\rmm{H},
\end{equation}
where $\sigma_\rmm{H}$ of Eq.~\ref{eq:maxwell0} is the charge of the Helmholtz plane on the right of the cell in  Fig~\ref{fig:stern1}, or equivalently minus the net charge on the Helmholtz plane on the left. Note that the sign of the polarization follows the convention for the electric field which changes sign from one layer to the next as indicated by the arrows in Fig~\ref{fig:stern1}. 

Polarization of the Stern layer will also add to, or subtract from, the fixed charge $\pm \sigma_0$  on the crystal planes. Similar to Eq.~\ref{eq:maxwell0} the net charge density of the terminal crystal plane would therefore be the sum of the fixed charge density $\sigma_0$ and the charge density induced at the bounding surface of the Helmholtz layer. However, consistent with the derivation in  Ref.~\citenum{Zhang2016b} we assume again that the gaps between the crystal planes are filled up by dielectric material.  The polarization in the gaps between the crustal planes is finite. Labelling the outer plane on the left by index $i=11$ we have for its net charge density       
\begin{equation}
\sigma_1 = \sigma_0 - P_\rmm{H} - P_1,
\label{eq:maxwell1} 
\end{equation}
where $P_1$ is the polarization between the outer plane 1 and plane 2, the next plane in from plane 1. With the convention for signs and directions of Fig.~\ref{fig:stern1} Eq.~\ref{eq:maxwell1} gives the charge on the left surface plane (blue dots).      

While the dielectric spacers consist of the same material everywhere, the polarization inside the crystal is not uniform. As a result, the polarization $P_2$ in the layer between plane $i=2$ and the next plane in (plane $i=3$) can therefore be different from $P_1$.  The negative charge of the first sub surface plane from the left (yellow dots) is therefore     
\begin{equation}
\sigma_2 = \sigma_0 - P_1 - P_2 .
\label{eq:maxwell2}
\end{equation}
As the fixed charge of the crystal planes alternates between $+\sigma_0$ and $-\sigma_0$ the Maxwell interface theorem for the series of planes $i=2 \dots n$ is satisfied by setting the charge to $\sigma_2$ switching sign from one plane to next ending with $-\sigma_1$ for plane $i=n+1$ on the right. 

The net surface charge densities of Eqs.~\ref{eq:maxwell0}-\ref{eq:maxwell2} are sufficient information to evaluate the polarization and Maxwell electric field of the continuum model of Fig.~\ref{fig:stern1}. The polarization is obtained form the total dipole moment  
\begin{equation} \label{eq:maxwell3}
P^\rmm{ICS}_\rmm{cell} = \frac{1}{L} \Big(\sigma_\rmm{H}(2l_\rmm{H} +nR) +
\sum_{i=1}^{n+1} \sigma_i z_i  \Big),  
\end{equation}
where $z_i$ is the location of the plane $i$ with surface charge density $\sigma_i$.  We have added a superscript ICS to indicate that the polarization of Eq.~\ref{eq:maxwell3} is specific for the insulator centered supercell of Fig.~\ref{fig:stern1}. The role of the subscript ``cell'' will become clear when we consider possible further contributions to the polarization from the electrostatic boundary conditions. Substitution of  Eqs.~\ref{eq:maxwell0}-\ref{eq:maxwell2} gives 
\begin{equation} \label{eq:Pdipole}
P^\rmm{ICS}_\rmm{cell} = \frac{1}{L} \Big(\sigma_\rmm{H}(2l_\rmm{H} +nR)-\sigma_1nR+\sigma_2\frac{n-1}{2}R\Big),
\end{equation}
Recall that $n$ is an odd integer. Note also that the polarization includes the dipole moment arising from the fixed charge density.

The Maxwell field $\bar{E}$ is defined as $ \Delta V = -L \bar{E} $ where $\Delta V$ is the potential across the cell. $\Delta V$ is obtained by adding all the potentials over the piece wise uniform partitions of the cell. Taking into account the convention for the sign of the fields specified in Fig.~\ref{fig:stern1} we find   
\begin{equation}
 L\bar{E} = - 2l_\rmm{H} E_\rmm{H} + \left(\frac{n+1}{2}\right) R E_1
-  \left(\frac{n-1}{2}\right) R E_2.
\label{eq:barE} 
\end{equation}
$\bar{E}$ can be finite leading to a finite potential over a periodic cell, which may at first seem to violate periodic boundary conditions. This is however a central feature of the SSV method avoiding open boundary conditions in condensed systems\cite{Stengel2009b}. 

The polarization of Eq.~\ref{eq:Pdipole} can be related to the Maxwell field $\bar{E}$  by replacing the surface charge densities by the  uniform fields in the subsystems  obtained employing the Maxwell interface equation at their boundaries 
\begin{eqnarray}
\label{eq:maxwell4}
4\pi\sigma_\rmm{H} &=& E_\rmm{H} \\
\label{eq:maxwell5} 
4\pi\sigma_1 &=& E_\rmm{H} + E_1 \\
4\pi\sigma_2 &=&  E_2 + E_1,
\label{eq:maxwell6}
\end{eqnarray}
where we have used that the field in the electrolyte is strictly zero in our simple model. Eqs.~\ref{eq:maxwell4}-\ref{eq:maxwell6} are equivalent to Eqs.~\ref{eq:maxwell0}-\ref{eq:maxwell2} formulated in terms of electric fields instead of polarization.  Inserting in Eq.~\ref{eq:Pdipole} gives
\begin{align}
4 \pi P^\rmm{ICS}_\rmm{cell}=~& \frac{1}{L} \Big(E_\rmm{H}(2l_\rmm{H}+nR)
 -(E_\rmm{H}+E_1)nR \nonumber
\\& +(E_1+E_2)\left(\frac{n-1}{2}\right)R\Big) 
\end{align}
Comparing to Eq.~\ref{eq:barE} we conclude that
\begin{equation}
4 \pi P^\rmm{ICS}_\rmm{cell}= -\bar{E}.
\label{eq:PICScell}
\end{equation}
Eq.~\ref{eq:PICScell} has the familiar form of the electric field generated by the polarization. At first glance this suggests that $\bar{E}$ is a polarization field. However, as defined by Eq.~\ref{eq:barE}, $\bar{E}$ is the Maxwell field including a possible contribution from an applied field (commonly referred to as $E_0$). In Ref.~\citenum{Zhang2016b} the resolution of this apparent inconsistency is that $ P^\rmm{ICS}_\rmm{cell}$ is not the  full polarization. What is missing is a surface term introduced by the supercell boundary. For the ICS this boundary is located in the electrolyte. The polarization in the electrolyte is uniform and therefore the same as the polarization at the boundary with the Stern layer.  This is the induction charge $-\sigma$ and hence we must write for the full polarization
\begin{equation}
\label{eq:PICS}
 P^\rmm{ICS}= -\sigma + P^\rmm{ICS}_\rmm{cell},
\end{equation}
or in terms of the Maxwell field using Eq.~\ref{eq:PICScell}
\begin{equation}
\label{eq:PICSE}
4 \pi  P^\rmm{ICS}= -4 \pi \sigma - \bar{E}.
\end{equation}
 
The need for surface terms related to periodic supercell boundaries is an important theme in solid state physics\cite{Martin1974}. In the physical chemistry literature on the simulation of polar liquids  this surface is often viewed as the polarization charge induced on the tinfoil  boundaries at infinity\cite{Zhang2016a,DeLeeuw1980a}.

The derivation to this point is a recapitulation of the argument in paper I. We now go beyond paper I by making the link to the dielectric displacement $D$.  As was pointed out, all charge in the system, including the fixed charge $\sigma_0$, is accounted for in the polarization Eq.~\ref{eq:Pdipole}. There is no external charge. The value of $D$ can therefore simply obtained by adding the polarization to the field according to Eq.~\ref{eq:defD}
\begin{equation}
\label{eq:DICS}
D^\rmm{ICS} = \bar{E}+ 4 \pi P^\rmm{ICS}.
\end{equation}
We have attached a superscript ICS anticipating that the dielectric displacement, like the polarization, depends on the choice of supercell. Substituting Eq.~\ref{eq:PICS} we find
\begin{equation}
  D^{\textrm{ICS}} = - 4 \pi \sigma.
  \label{eq:Dsigma}
\end{equation}
Compared to the complexity of the expression for the Maxwell field (Eq.~\ref{eq:barE}) the equation for $D$ is surprisingly simple, which is another reason for preferring the constant $D$ over the constant $E$ method. Note however that the surface charge density determining the displacement field in Eq.~\ref{eq:Dsigma} is the induced charge $\sigma$, not the fixed charge $\sigma_0$, which would have been the expected value if the fixed charge had been treated as external charge.

\subsubsection{Compensating field} \label{sec:ecnc}

The idea of paper I was to eliminate the finite field in the solid slab by imposing a biasing field $E_\rmm{CNC}$. This field was found by requiring that the potential over the crystal slab vanishes. Similar to Eq.~\ref{eq:barE} for the potential over the entire periodic cell, the potential over the crystal can be interpreted in terms of the average electric field
\begin{equation}
E_d = -\frac{n-1}{2n}E_2+\frac{n+1}{2n}E_1;
\label{eq:Ed}
\end{equation}
while the interior of the crystal is not a dielectric as in Ref~\citenum{Zhang2016b} we have kept the same subscript $d$. Setting $E_d =0$ gives a linear relation between $E_1$ and $E_2$. 
\begin{equation}
\label{eq:Ed0}
  (n-1)E_2 = (n+1)E_1.
\end{equation}
At this point we finally invoke the linear constitutive relations $4 \pi P_{\textrm{H}} =  ( \epsilon_{\textrm{H}}-1)  E_{\textrm{H}}$ for the polarization in the Stern layer and $ 4 \pi P_1 = (\epsilon_d-1) E_1, \; 4 \pi P_2 = (\epsilon_d-1) E_2$ for the polarization in the dielectric material between the crystal planes. Substituting in Eqs.~\ref{eq:maxwell0}--\ref{eq:maxwell2} and combining with Eqs.~\ref{eq:maxwell4}-\ref{eq:maxwell6} we have
\begin{eqnarray}
\epsilon_\rmm{H}E_\rmm{H} &=& 4\pi\sigma \label{eq:maxwell7}\\ 
\epsilon_\rmm{H}E_\rmm{H} + \epsilon_dE_1 &=& 4\pi\sigma_0 \label{eq:maxwell8}\\ 
\epsilon_dE_2 + \epsilon_dE_1 &=& 4\pi\sigma_0, \label{eq:maxwell9}
\end{eqnarray}
Eq.~\ref{eq:maxwell9} together with Eq.~\ref{eq:Ed0} gives 
\begin{numcases}{\rmm{CNC}\Rightarrow}
E_2 = \frac{4\pi}{\epsilon_d}\left(\frac{n+1}{2n}\right)
\sigma_0 \label{eq:E2CNC} \\[4pt]
E_1 = \frac{4\pi}{\epsilon_d}\left(\frac{n-1}{2n}\right) \sigma_0 \label{eq:E1CNC}
\end{numcases}
Replacing $E_1$ in Eq.~\ref{eq:maxwell8} with Eq.~\ref{eq:E1CNC} and then substituting in Eq.~\ref{eq:maxwell7} we recover Eq.~\ref{eq:sigman}. Inserting Eqs.~\ref{eq:E2CNC} and \ref{eq:E1CNC} into Eq.~\ref{eq:barE} we find our expression for the compensating field
\begin{equation}
\label{eq:ECNC}
E_\rmm{CNC} = - \frac{4\pi}{\epsilon_\rmm{H}} 
 \left(\frac{l_\rmm{H}}{L}\right) \left(\frac{n+1}{n}\right) \sigma_0.
\end{equation}
Comparing Eq.~\ref{eq:ECNC} for the compensating field to Eq.~\ref{eq:sigman} for $\sigma$ under CNC conditions ($E_d =0$) we note that all parameters of the continuum model except the number of planes ($n$) have disappeared.  This would suggest that the validity of Eq.~\ref{eq:sigman} extends beyond he continuum model as was confirmed by the close agreement found for the FFMD model in paper I.

Eq.~\ref{eq:Dsigma} is generally valid whether under CNC conditions or not. When $\bar{E}=E_\rmm{CNC}$ the induced charge is determined by the fixed charge $\sigma_0$ according to Eq.~\ref{eq:sigman}.  Inserting this equation in Eq.~\ref{eq:Dsigma} yields the value of the displacement field at CNC.
\begin{equation}
 \label{eq:DCNC}
D_\rmm{CNC}^\rmm{ICS} = -4\pi \left(\frac{n+1}{2n} \right) \sigma_0,
\end{equation}
As hypothesised, the value of $D_{\rmm{CNC}}$ is known \textit{a priori}. The consequence of this is that CNC does not have to be located by scanning over E~fields and performing an interpolation. This is crucial, as in our previous work this amounted to a minimum of 5~separate trajectories. Polarization in these systems approaches a value comparable to convergence after 100-200~ps, and while this is trivial for a classical code, our ambition is to include electronic structure, which will in general produce a significantly different value of $P_\rmm{CNC}$. This means the search over E~fields would have to be repeated using DFTMD. In contrast, the model value of $D_\rmm{CNC}$ will be the same in both FF and DFT descriptions.


\subsubsection{Equation of State} \label{sec:eos}

In the EDL study of Ref.~\citenum{Zhang2016b} we observed that the response of the polarization to a finite field was remarkably linear, even for relatively large fields. This suggested writing the electric equation of state in linear form 
\begin{equation} \label{eq:Estate}
4\pi P^\rmm{ICS}= 4 \pi \gamma_E\sigma_0+({\epsilon}_\perp-1)\bar{E},
\end{equation} 
where $\gamma_\rmm{E}$ and ${\epsilon}_\perp$ are constants. The interpretation of ${\epsilon}_\perp$ is as a `global' dielectric constant for the composite system. It was shown that the capacitance could be estimated without finding CNC by calculating the derivative of the potential with respect to the surface charge, opening a complimentary route to calculation of this important observable.

Here, the expression for the charge in the double layer is subjected to the same analysis. We first rewrite the expression for the Maxwell field (Eq.~\ref{eq:barE}) as the sum of the potential over the two Stern layers (assumed identical in our model) and the potential over the crystal (Eq.~\ref{eq:Ed})
\begin{equation}
\label{eq:es1}
\bar{E}L=-2E_\rmm{H}l_\rmm{H}+n\bar{E}_\rmm{d}R,
\end{equation} 
 With some manipulation of Eq.~\ref{eq:Ed} and Eqs.~\ref{eq:maxwell7}-\ref{eq:maxwell9}, $E_\rmm{d}$ can be expressed as 
\begin{equation}
\label{eq:es2}
 E_d = - \frac{4 \pi}{\epsilon_d} \left( \sigma-
 \frac{n+1}{2n} \sigma_0 \right),
\end{equation}
the difference between $\sigma$ and its value under CNC bias (Eq.~\ref{eq:sigman}). This is as expected because the average field in the crystal vanishes at CNC. Substituting in Eq.~\ref{eq:es1} gives $\sigma$ in the form~\footnote{This is Eq. 6 of Paper I, which contained a typo.}
\begin{equation} 
\label{eq:sigma}
\sigma = \left(\frac{n+1}{2}\frac{\sigma_0}{C_d}-\bar{E}L\right)\left(\frac{2}{C_\rmm{H}}+\frac{n}{C_d}\right)^{-1},
\end{equation}
where $C_\rmm{d}=\epsilon_\rmm{d}/(4\pi R)$ and $C_\rmm{H}=\epsilon_\rmm{H}/(4\pi l_\rmm{H})$. Substituting in expression Eq.~\ref{eq:PICSE} for the full polarization we find for the parameters of the equation of state Eq.~\ref{eq:Estate} 
\begin{eqnarray}
\label{eq:bareps}
{\epsilon}_\perp  & = & 4\pi L C_\text{tot},
\\
\label{eq:gammaE1}
\gamma_E & = & -  \left( \frac{n+1}{2}\right)\frac{C_\text{tot}}{C_d}.
\end{eqnarray}
${C}_\text{tot}$ is the series capacitance of polar slab including the two double layers at either end and defined as:
\begin{equation}
\label{eq:barC}
\frac{1}{{C}_\text{tot}} = \frac{2}{C_\rmm{H}}+\frac{n}{C_d}.
\end{equation}

Inspecting the large $n$ (thick slab) behaviour of the equation of state reveals some surprising features. In this limit the contribution $2/C_\rmm{H}$ of the Helmholtz layer to series capacitance ${C}_\text{tot}$ (Eq.~\ref{eq:barC}) can be neglected. This reduces the composite dielectric constant Eq.~\ref{eq:bareps} to  $\epsilon_\perp = (L/nR)\epsilon_d$. This would imply that the value $\epsilon_\perp$ could decrease below unity, which is a forbidden interval for dielectric constants. However $L=nR + 2 l_\rmm{H}$ coupling box length and slab width. Rigorously, using the definitions of $C_\rmm{H}$ and $C_d$ 
\begin{equation}
\frac{1}{4 \pi {C}_\text{tot}}
=  \frac{ 2 l_\rmm{H}}{\epsilon_\rmm{H}} + \frac{ nR}{\epsilon_d} <  
2 l_\rmm{H} + nR < L
\end{equation}
which guarantees that ${\epsilon}_\perp > 1$. To investigate the large $n$ limit of $\gamma_E$ we write Eq.~\ref{eq:gammaE1} as 
\begin{equation}
\label{eq:gammaE2}
\gamma_E \sigma_0 = - \frac{n C_\text{tot}}{C_d} \sigma_{\rm{CNC}} 
\end{equation}
where we have made use of Eq.~\ref{eq:sigman}. For large $n$ the prefactor
approaches unity leading to $\gamma_E \sigma_0 =  - \sigma_{\rmm{CNC}}$. This is rather surprising, because Eq.~\ref{eq:sigman} is derived by imposing CNC conditions, while the equation of state Eq.~\ref{eq:es1} is generally valid (given the linear response approximation of the continuum model). Evidently the zero field polarization  $(\bar{E}=0$) converges to the CNC value for increasing slab width.   

The main objective of our study of the EDL in Ref.~\citenum{Zhang2016b} was the computation of the capacitance $C_{\rmm{H}}$ of the Stern layer. This quantity is can also defined for the polar surface as can be seen by rewriting E.~\ref{eq:ECNC} as 
\begin{equation}
\label{eq:VCNC}
 \Delta V_{\rmm{CNC}} = -L E_{\rmm{CNC}} = \frac{2}{C_{\rmm{H}}} \sigma_{\rmm{CNC}}
\end{equation} 
$\Delta V_{\rmm{CNC}}$ is the potential over the periodic cell. The  potential difference over the crystal vanishes at CNC. The potential over the electrolyte is always zero, and therefore $\Delta V_{\rmm{CNC}}$ is the sum of the potentials over the compact double layers. Recall however that for polar surface the double carries a net charge and Eq.~\ref{eq:VCNC} must be regarded a formal definition of the double capacitance.      

$C_{\rmm{H}}$ can be estimated directly from Eq.~\ref{eq:VCNC}.  As an alternative we considered in  Ref.~\citenum{Zhang2016b}  to estimate the compact layer capacitance from the response to the variation in $\bar{D}$ given by the conjugate form of equation~\ref{eq:Estate},
\begin{equation} \label{eq:Dstate}
4\pi P^\rmm{ICS}= 4 \pi \gamma_\rmm{D}\sigma_0+\left(1-\frac{1}{\epsilon_\perp}\right)\bar{D}.
\end{equation}
where $\gamma_D = \gamma_E/\epsilon_\perp$. This is significant because the aqueous response to the $\bar{D}$~field is known to converge faster than its conjugate by a factor equal to its dielectric constant. However, depending on the actual form of the estimator, the gain in the convergence of the polarization does not always lead to a speed-up of the calculation of dielectric properties and a good example is the direct application of constant $D$ simulations to compute the dielectric constant of liquid water~\cite{Zhang2016a}. We will come back to this issue in Sec.~\ref{sec:results}.  


A third option for computing the double layer capacitance is from the value of polarization at CNC. Writing the polarization at CNC as the difference between the corresponding values of the displacement field (Eq.~\ref{eq:DCNC}) and the Maxwell (Eq.~\ref{eq:VCNC}) we can find  
\begin{equation} 
\label{eq:capacitance}
C_{\rmm{H}}=\frac{\sigma_{\rmm{CNC}}}{2\pi L}
 \left( \sigma_{\rmm{CNC}}+P^\rmm{ICS}_{\rm{CNC}} \right)^{-1}.
\end{equation}
The results for the capacitance of the double layer reported later were calculated using Eq.~\ref{eq:capacitance} although also suffering from deterioration in accuracy in the limit of large width.  $P^\rmm{ICS}_{\rm{CNC}}$ approaches $-\sigma_{\rm{CNC}}$ as $1/L$, keeping $C_{\rmm{H}}$ finite, but with increasing statistical error.  


\subsection{Electrolyte centered supercell (ECS)} \label{sec:ecs} 

In an electrolyte centered supercell (ECS) the electrolyte is fully contained in the cell. The boundaries of the cell are now cutting through the crystal. In the present exploratory investigation the position of the ions in the crystal are still kept fixed. This leaves only two options of where to cut. Referring to the definitions of Fig.~\ref{fig:stern1}, the boundaries of the cell can either be located in an interplanar layer with electric field $E_1$ or in a layer with electric field $E_2$. We start with the $E_1$ which the easier to understand. Also to begin, one cell boundary is placed just below the surface plane in the first $E_1$ layer. With one side of the supercell contained only one crystal plane, the other side must contained the remaining $n$ ion planes. The polarization of this cell (total dipole moment divided by boxlength $L$) is now computed as  
 \begin{equation}
\label{eq:PECS1a}
P_\rmm{cell}^\rmm{ECS1}
 = \frac{1}{L} \Big(-\sigma_\rmm{H}l_e +\sigma_1(2l_\rmm{H}+l_e)+\sigma_2\frac{n-1}{2}R\Big) 
\end{equation}
where we have augmented the supercell superindex to indicate the type of cut. 
$l_e$ is the length of the electrolyte region and therefore
\begin{equation}
\label{eq:lsum}
 L= l_e + 2 l_\rmm{H} + nR.
\end{equation}
 The interface equations for the net charges are independent of cell geometry, so we can still use Eqs.~\ref{eq:maxwell4}-\ref{eq:maxwell6}. Substituting gives
\begin{eqnarray}
4 \pi P_\rmm{cell}^\rmm{ECS1}
&=& \frac{1}{L} \Big(E_\rmm{H}2l_\rmm{H}+
\nonumber\\[4pt] &&
E_1 L-E_1 nR+(E_1+E_2)\frac{n-1}{2}R\Big) 
\end{eqnarray}
where we have eliminated $l_e$ using the geometric relation Eq.~\ref{eq:lsum}.
The Maxwell field $\bar{E}$ consists of a sum of potentials 
and is not affected either by a change of cell boundaries ($\bar{E}$ must be the same because it acts as a force on the particles and is therefore an observable). Thus using Eq.~\ref{eq:barE} we can write
\begin{equation}
\label{eq:PECS1b}
4 \pi P_\rmm{cell}^\rmm{ECS1} =E_1-\bar{E},
\end{equation}
which has an additional term of $E_1$ on the right-hand side when compared with the ICS version of Eq~\ref{eq:PICScell}. 

As with the ICS, the ECS polarization is only complete after adding the appropriate cell boundary surface term. The field in the dielectric intersected by the boundary is $E_1$, the surface term is therefore $P_1$.   Adding to Eq.~\ref{eq:PECS1b} and combining with the $E_1$ term using constitutive relation yields
\begin{equation}
\label{eq:PECS1c}
4 \pi P^\rmm{ECS1} =\epsilon_d E_1-\bar{E}
\end{equation}
This expression was derived for a specific choice of $E_1$ cut. One of the two sections of the crystal in the cell consists only of one ionic plane, a surface plane. A more evenly dividing  $E_1$ cut can be generated by moving over pairs of neighbour planes leaving the outer planes in place.  Such a pair of planes is charge neutral. Translating the pair will not alter the total dipole moment and Eq.~\ref{eq:PECS1c} remains valid. 

To find the expression of polarization for an $E_2$ cut we start again placing one cell boundary as near as possible to a surface plane. For an $E_2$ cut that is the second subsurface layer between the second and third plane of ions. The total dipole moment Eq.~\ref{eq:PECS1a} is modified to 
 \begin{eqnarray}
\label{eq:PECS2a}
P_\rmm{cell}^\rmm{ECS2}
 & = &  \frac{1}{L} \Big(-\sigma_\rmm{H}l_e+\sigma_1(2l_\rmm{H}+l_e)+
 \nonumber \\ [4pt]
 && - \sigma_2 (2R + 2l_\rmm{H}+l_e) - \sigma_2\frac{n-3}{2}R\Big) 
\nonumber \\ [4pt]
 & = &  \frac{1}{L} \Big(-\sigma_\rmm{H}l_e+\sigma_1(2l_\rmm{H}+l_e)+
 \nonumber \\ [4pt]
&& -\sigma_2 L + \sigma_2 \frac{n-1}{2} R \Big)
\end{eqnarray}
Note that the direction of the dipole of a pair of the planes beyond the first two planes in Eq.~\ref{eq:PECS2a} is pointing in the opposite direction   relative to the dipole of paired planes in Eq.~\ref{eq:PECS1a}. Comparing to Eq.~\ref{eq:PECS1a} we see there is now an extra contribution $-\sigma_2$. 
 \begin{equation}
 P^\rmm{ECS2}_\rmm{cell} =  P^\rmm{ECS1}_\rmm{cell} -\sigma_2 
\end{equation}
The cell polarization of the $E_2$ cut can therefore be immediately obtained from Eqs.~\ref{eq:PECS1b} using again Eq.~\ref{eq:maxwell6}
\begin{equation}
 \label{eq:PECS2b}
4 \pi P^\rmm{ECS2}_\rmm{cell}
 = -E_2 - \bar{E}.
\end{equation}
Perhaps not very surprising, the cell specific offset of $\bar{E}$ has changed from the field in the layer intersected by the $E_1$ cut to the corresponding field for an $E_2$ cut (note the opposite sign is a result of the convention of the field directions in Fig.~\ref{fig:stern1}). The same applies to the surface term which is now given by $-P_2$ leading in total to a cell polarization of     
\begin{equation}
 \label{eq:PECS2c}
4 \pi P^\rmm{ECS2} = - \epsilon_d E_2 - \bar{E} .
\end{equation}
Again, paired planes can be moved to the other side. Eq.~\ref{eq:PECS2c} is therefore the general expression for an $E_2$ cut. Subtracting the $E_1$ polarization Eq.~\ref{eq:PECS1c} we find
\begin{equation}
\label{eq:PECS2d}
  P^\rmm{ECS2} - P^\rmm{ECS1}  = - \frac{\epsilon_d}{4 \pi} \left(E_1 + E_2 \right) = -\sigma_0 
\end{equation}
where in the second step we have substituted Eq.~\ref{eq:maxwell9}. This is to be expected, we have moved effectively one plane more to generate an $E_2$ cut. 
 
How to relate the ECS to ICS polarization? The $E_2$ cut turns out to lead the same polarization as for an ICS geometry. This can be seen by subtracting Eq.~\ref{eq:maxwell9} from Eq.~\ref{eq:maxwell8}, eliminating $E_1$ and giving $\epsilon_d E_2 = \epsilon_\rmm{H} E_\rmm{H}$ which then via Eq.~\ref{eq:maxwell7} becomes  $\epsilon_d E_2 = 4 \pi \sigma$. Inserting in Eq.~\ref{eq:PECS2c} we obtain
\begin{equation}
 \label{eq:PECS2e}
4 \pi P^\rmm{ECS2} = - 4 \pi \sigma - \bar{E}.
\end{equation}
We indeed recover the same polarization as the ICS Eq.~\ref{eq:PICSE}. This can be rationalized by counting the number of planes crossing the cell boundary when the supercell is shifted from the ICS to an ECS geometry. Every time a crystal plane leaves the cell and renters on the other side the cell polarization jumps by $\pm \sigma_2$.  The surface term alternates in step between its two values.  For the $E_2$ variety of ECS cell the numbers of jumps is even cancelling each other. For the $E_1$ cut the number of planes crossing the boundary is odd. The effective plane charge and boundary charge do not cancel but add to a net charge of $\pm \sigma_0$ depending on whether the shift is to the left or right. The $E_1$ cut could therefore also be referred to as an ``odd'' ECS (ECS1) and the `$E_2$ as an ``even'' ECS (ECS2). In the following we prefer the more appealing odd/even terminology over the somewhat awkward $E_1$ cut/$E_2$ cut method of referencing. 

The displacement field $D$ is obtained by adding the invariant Maxwell field. $D$ will therefore exhibit the same changes as $P$. In particular the $D$ field at CNC in the $E_2$ ECS is given by Eq.~\ref{eq:DCNC}. For the $E_1$ cut this value must be corrected by $\sigma_0$ as required by Eq.~\ref{eq:PECS2d}  
\begin{equation} \label{eq:ourD}
D^\rmm{ECS1}_\rmm{CNC}= 4\pi \frac{n-1}{2n}\sigma_0.
\end{equation}
The important conclusion of this section is that the displacement field values imposing CNC conditions, while derived from a continuum model, are in the end independent of the dielectric parameters $\epsilon_d, \epsilon_\rmm{H}$ and even of the geometric parameters $R, l_\rmm{H}$. The only parameter that matters is the bare charge density of the ionic planes. This gives us confidence that the relation for $D$ are generic and can be transferred to our atomistic models. At present however we are not able to give a convincing ``model free'' explanation for this observation.    

\section{Electronic Polarization}\label{sec:aimd}

 
The modern theory of polarization was developed by Vanderbilt, Resta and colleagues to describe and compute electronic polarization in insulating solids (semiconductors, ionic crystals) in the framework of the Density Functional Theory (DFT) based  band structure calculation in periodic supercells \cite{King-Smith:1993prb,Resta:1994rmp,Resta:2007ch}. As had been realized already for some time (see for example Martin\cite{Martin1974}) polarization in such systems cannot be computed from the electronic density but must be treated as an independent system variable. The breakthrough came when this  variable was identified by Vanderbilt and coworkers as a Berry phase related to the Bloch orbitals in reciprocal space\cite{King-Smith:1993prb}.  Resta subsequently reformulated the Berry phase for polarization in terms of a phase in real space\cite{Resta:1998prl,Resta:2000jpcm}. The implementation of Berry-Phase electronic polarization in CP2K code employed in the present calculation is based on the $\Gamma$-point only Resta approach\cite{Hutter:2013iea}. An alternative real approach for the calculation of electronic polarization uses Maximally Localized Wannier Functions (MLWFs)~\cite{Marzari:1997wa,Silvestrelli:1998ssc,Marzari:2012eu}. The polarization is obtained from the dipole moment of the centers of the MLWFs. Polarization in this definition is again multivalued because of the freedom in choice which periodic image to use in common with the classical itinerant polarization of Sec.~\ref{sec:spcpol}.       

This poses a practical problem of branch alignment. The $ n \Delta_q P$ gauge of the polarization upon initialization of the electron structure calculation is not easy to control.  However, as pointed out in Sec.~\ref{sec:spcpol} it is crucial that the gauge of the displacement field in the constant D Hamiltonian matches the gauge of the polarization. The challenge is now: how to map the Berry phase polarization being handled by the computer code to the value we recognize as being consistent with our model? For this we consider our initial geometry in the basis of the Wannier-center representation of the electronic wavefunction. The localized electron pairs are attached to nearby atoms and these atoms are wrapped into the box as whole molecules. This `molecular gauge' gives us the anchoring value of the polarization, and differs from other polarization branches by $n \Delta_q P$ for integer $n$. 

In principle we could then go ahead with the ICS geometry. However, since solvent molecules are not defined beforehand -- as in the case of FFMD simulation -- and are free to straddle across the supercell boundary or (more seriously) dissociate, we would have to keep track of a frequently changing polarization branch. This can be avoided by conveniently shifting one-half box length to the ECS. Now, no aqueous species can cross the boundary, and since the crystal will be held fixed, neither will any part of the lattice. The only change of branch will be from the electrons within the crystal crossing the boundary to localize on a different plane of ions.

\begin{figure}[h]
\begin{center}
\includegraphics[scale=0.28]{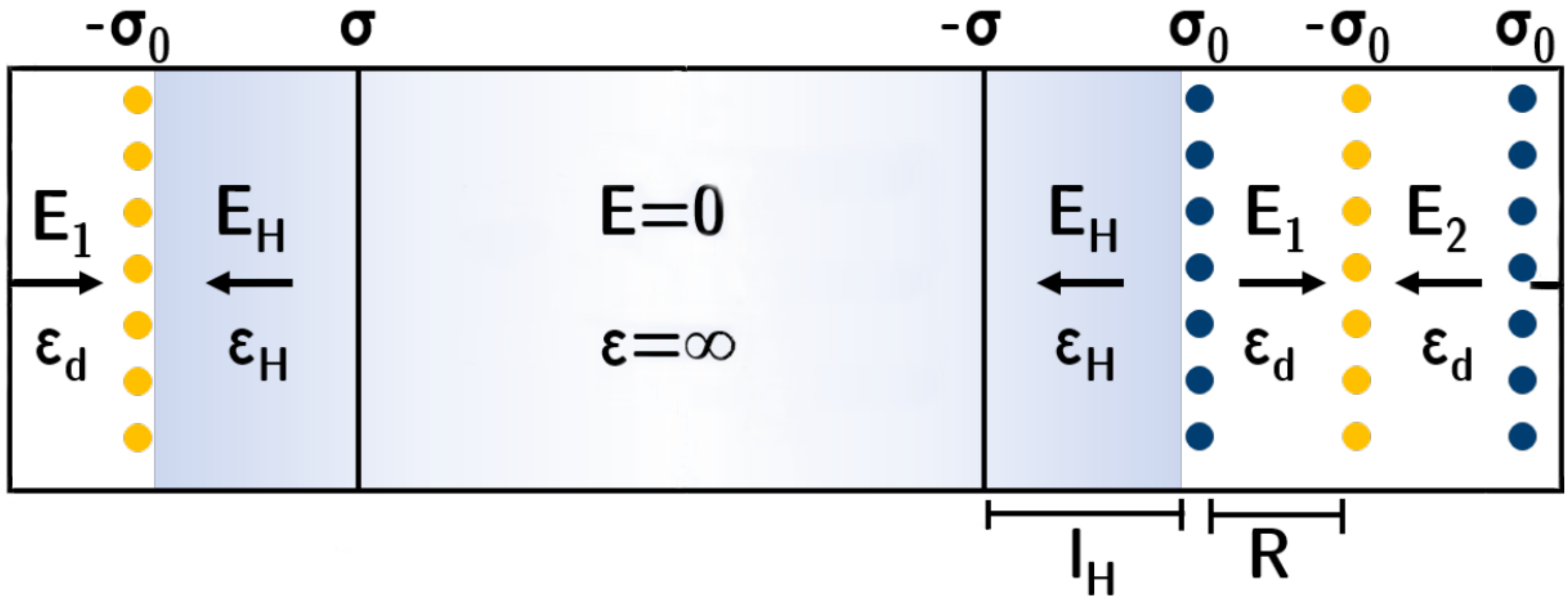}

\includegraphics[scale=.725]{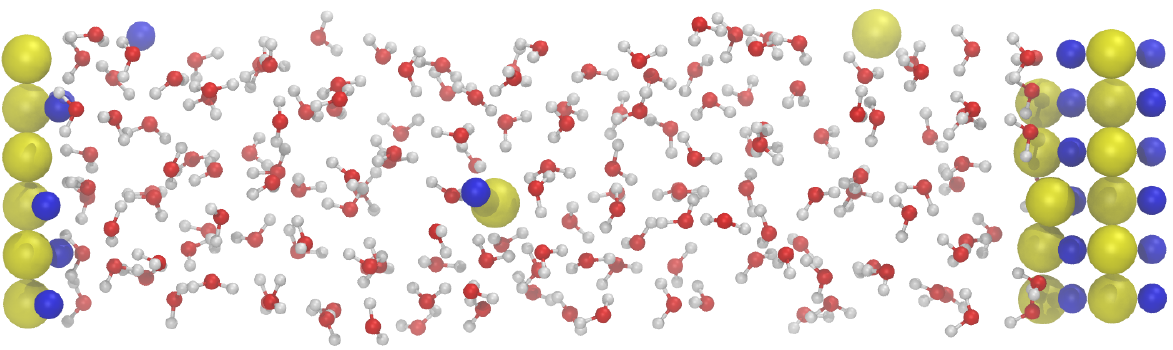}
\end{center}
\caption{Above: Stern model of the ECS. The lone plane on the left hand side means that $E_1$ is cut by the supercell boundary and so this is on an `odd' polarization branch ($P^\text{ECS1}$). All the symbols have the same meaning as in Fig.~\ref{fig:stern1}. Below: MD snapshot of the NaCl(111)/NaCl solution system, with Na$^+$ in blue and Cl$^{-}$ in yellow.} \label{fig:stern2}
\end{figure}

In this study we will use a DFTMD crystal slab of only $n=3$ and so $(n+1)/2$ is even and we could draw the cell boundary in the central $E_2$ layer. As argued in Sec.~\ref{sec:ecs} the displacement field  $D$ is the same for this ``even'' type  ECS (ECS1) and the ICS. But in order to demonstrate this supercell dependence we shall shift one plane to achieve an `odd' branch of the polarisation ($P^{\text{ECS1}}$) as shown in Fig.~\ref{fig:stern2}) and instead apply the smaller, positive D of Eq.~\ref{eq:ourD}.

\section{Computational Details} \label{sec:tech}

From our previous work we found that the model agreed remarkably well with the results from FFMD with as few as 4 planes of ions ($n=3$). We therefore feel confident to use this size of system in the DFTMD. Furthermore, we reduced the lateral extent of the cell by reducing the number of ions per plane (The $N_0$ in Eq.~\ref{eq:DqP0}) from 16 to 9. This also allows us to reduce to 7 aqueous ion pairs (previously 20) and still retain an electrolyte region. This is because the compensating charge is $A\sigma_0(n+1)/2n=6~\rmm{e}$. We can now reduce the $z$ axis without significantly affecting the initial concentration. Note that since the hexagonal symmetry of the (111) plane requires one of the dimensions to be an even number of planes, this manoeuvre introduced a stacking fault along the $x$ axis. This leads to a system of 539 atoms within a box of lengths [1.197,~1.0365,~4.75]~nm. We recomputed the capacitance for this smaller size to be 8.5~$\upmu$Fcm$^{-2}$, a minor increase of 3\% from the previous value of 8.2~$\upmu$Fcm$^{-2}$ for the system without the stacking fault in Paper I. We then tested doubling the lateral extent and this yields 8.4~$\upmu$Fcm$^{-2}$.  This deviation is small enough to justify our choice of minimalistic supercell. All FFMD simulations were performed under ambient conditions using a modified version of the GROMACS 4 package~\cite{Hess2008}. The water model was Simple Point Charge/Extended (SPC/E)~\cite{Berendsen:1987uu}, with the Na$^{+}$ and Cl$^{-}$ parameters of Joung and Cheatham~\cite{Joung2008} which has been validated for high ionic strength~\cite{Zhang:2010zh, Nezbeda2016}. Technical settings of the simulations were the same as used in paper I. In the analysis of the trajectories, the first 200~ps were discarded unless otherwise specified. 

The DFTMD simulations were performed with CP2K~\cite{VandeVondele:2005ge,Hutter:2013iea}. The pseudo-potentials used were Goedecker-Teter-Hutter (GTH)~\cite{Goedecker:1996ve}, with the double-zeta polarized DZVP-MOLOPT-SR-GTH basis set~\cite{VandeVondele:2007wt}, such that the nuclei of Na, Cl, O, and H have apparent charges of 9e, 7e, 6e, and 1e respectively. The exchange correlation functional was Perdew-Burke-Ernzerhof (PBE)~\cite{PhysRevLett.77.3865}, the timestep was 0.5~fs, the Bussi-Parrinello thermostat was set to be 298~K with a time constant of 20~fs~\cite{Bussi:2008wu}, Orbital Transformation (OT) was used with full single inverse with default (-1) stepsize and energy gap, convergence was 5E-7~\cite{VandeVondele:2003ue}. A charge cutoff of 320 Ry with 40 Ry for the relative grid was found to be sufficient.  The constant $D$ implementation in CP2K can be referenced to Ref.~\citenum{Zhang2016c} and is publicly available. 

\section{Results and Discussion} \label{sec:results}
\subsection{FFMD Validation}

\begin{figure}[h]
\begin{center}
\resizebox{.48\textwidth}{!}{\includegraphics{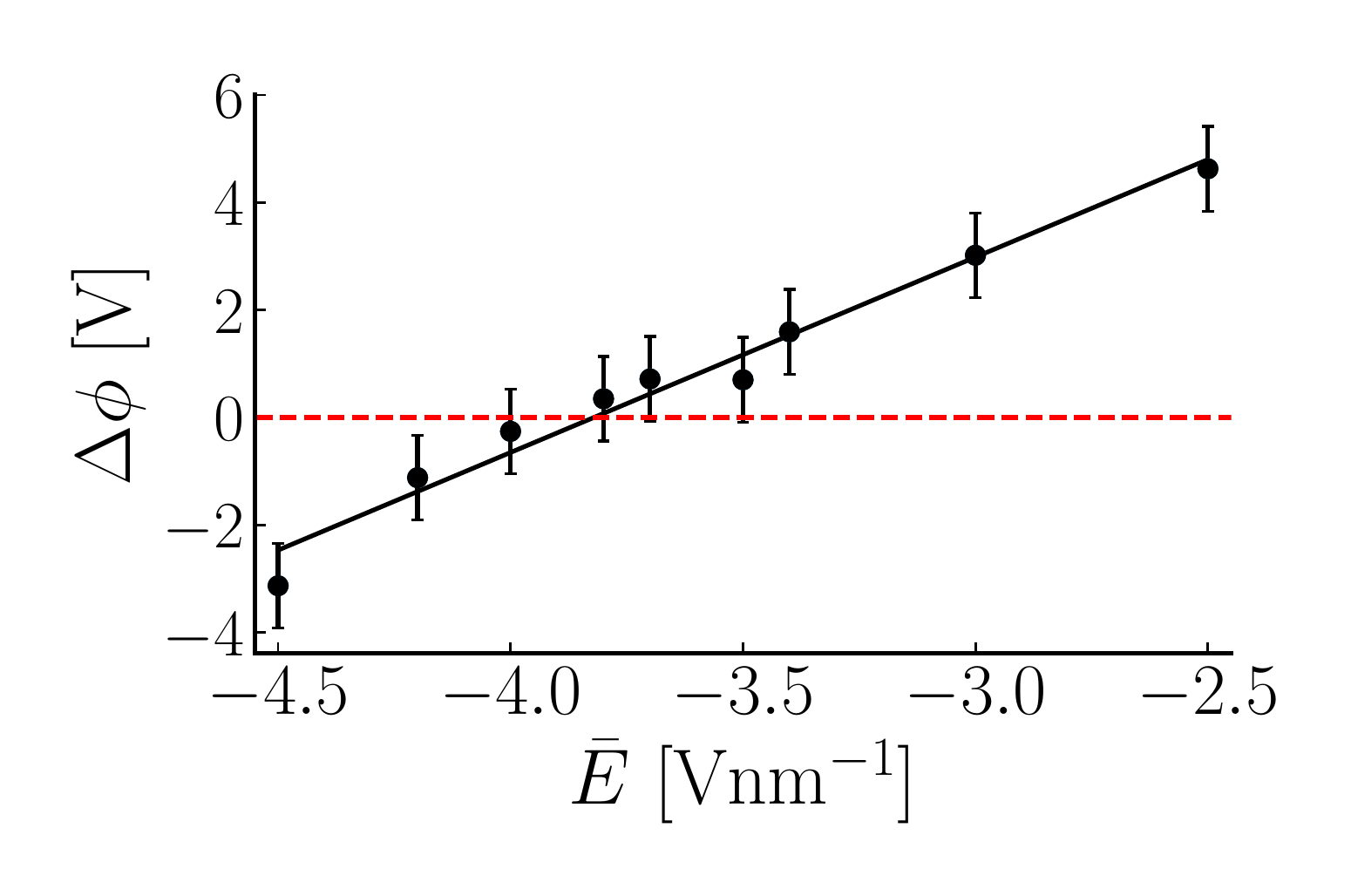}}
\end{center}
\vspace{-25pt}
\caption{FFMD simulations of the $n=3$ `odd' ECS (ECS1) of Fig~\ref{fig:stern2}. $E_\rmm{CNC}$ was found to be $-3.82~\rmm{Vnm}^{-1}$. Error bars are 3$\sigma$ obtained by jackknife resampling of the data.} \label{fig:efieldscan}
\end{figure}

The small system to be submitted to DFTMD was first studied with FFMD in order to verify the derived value of D$_\rmm{CNC}$ (Eq.~\ref{eq:ourD}). As in the previous work, a scan over E fields was performed, and an interpolation of the change in potential over the crystal found $E_\rmm{CNC}$ to be -3.82~Vnm$^{-1}$, as shown in Fig.~\ref{fig:efieldscan}.

\begin{figure}[h]
\begin{center}
\resizebox{.48\textwidth}{!}{\includegraphics{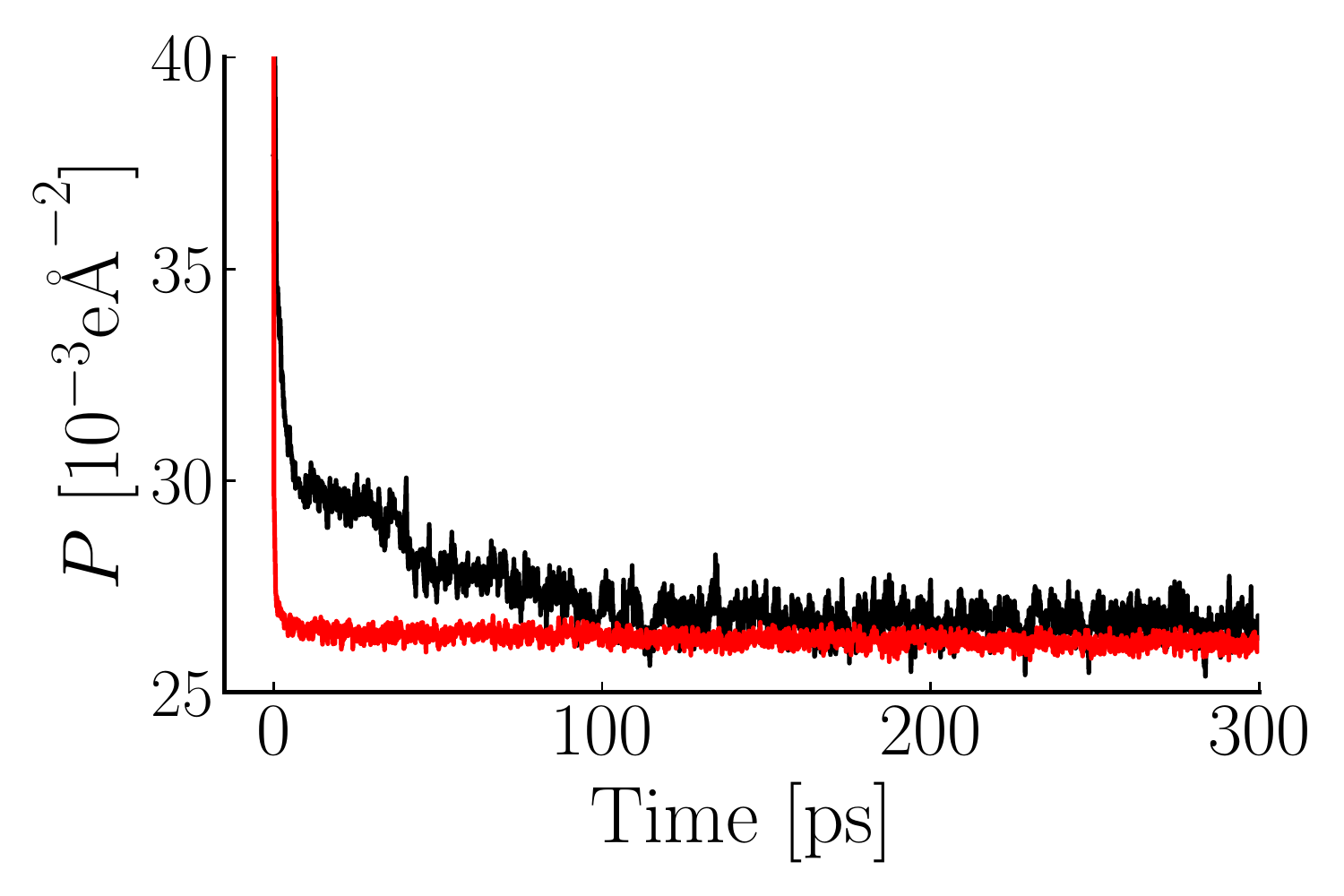}}
\end{center}
\vspace{-15pt}
\caption{FFMD simulations of the $n=3$ `odd' ECS (ECS1) of Fig~\ref{fig:stern2}. The red line is $D=43.75~\rmm{Vnm}^{-1}$, while the black line is $\bar{E}=-3.8~\rmm{Vnm}^{-1}$. Only the first 300~ps are displayed to emphasise the behaviour in the range 0-100~ps, the final values of $P$ are given in the text.} \label{fig:convergeclassic}
\end{figure}

For this system, $D^\rmm{ECS1}_\rmm{CNC}=43.75~\rmm{Vnm}^{-1}=24.18\times10^{-3}\rmm{e\AA}^{-2}$. Therefore we expected the polarization at CNC to be $26.34\times10^{-3}\rmm{e\AA}^{-2}$. Calculating the polarization as simply the sum of the classical charges multiplied by their positions (per unit volume) gave the slightly larger $26.78\times10^{-3}\rmm{e\AA}^{-2}$ for the point at 3.8~Vnm$^{-1}$, showing that our theoretical value of $D_\rmm{CNC}$ was accurate. We then ran the same CNC simulation but at a constant displacement field. We found $P_\rmm{CNC}=26.26\times10^{-3}\rmm{e\AA}^{-2}$ with a corresponding $E_\rmm{CNC}=-3.76 ~\rmm{Vnm}^{-1}$. The comparison is shown in Fig.~\ref{fig:convergeclassic}.

Therefore, the results of constant $D$ approach is consistent with that of the $E$~field interpolation and the difference was of the order $0.05~\rmm{Vnm}^{-1}$. This validates Eq.~\ref{eq:ourD} and the associated Eq.~\ref{eq:DCNC}. Further, it can be clearly seen in Fig.~\ref{fig:convergeclassic} that the $D_\text{CNC}$ field ensemble achieves a much faster convergence of the polarization over its thermodynamic conjugate condition $E_\text{CNC}$.

\subsection{DFTMD Simulations}

\subsubsection{Initialization and polarization alignment}
We carried over the last frame of FFMD trajectory to be the first frame of DFTMD simulations, after a short geometry optimization. Because of the multi-valued nature of polarization, the first task was to align the starting value of $P$ calculated from the Berry phase formalism to the anchoring polarization calculated from the maximally localized Wannier functions (MLWFs)~\cite{Marzari:1997wa,Marzari:2012eu} as explained in Sec.~\ref{sec:aimd}. The two values of polarization can differ by a multiple of the quantum of polarization $\Delta_qP$. Using Eq.~\ref{eq:DqP0} we computed a $\Delta_q P = 8.06\times 10^{-3}\rmm{e\AA}^{-2}$ for our setup. A test calculation at $D=0$ showed that CP2K had calculated the starting polarization as $1.42\times10^{-3}\rmm{e\AA}^{-2}$. By aligning this value to the molecular gauge obtained from MLWFs which is $25.61\times10^{-3}\rmm{e\AA}^{-2}$, we found out the starting value of $P$ in our DFTMD system differs by $3 \Delta_q P$. 

$D_\text{CNC}$ depends only on $\sigma_0$ and the number of crystal planes as indicated by Eq.~\ref{eq:ourD}. For $n=3$, $D_\text{CNC}= 4\pi \sigma_0/3 = 12 \pi \Delta_q P$ where $\sigma_0 = N_0 \Delta_q P = 9 \Delta_q P$ for our small system. This value is the same for both FFMD and DFTMD simulations in our setup. As pointed out already in Ref.~\citenum{Zhang2016b}, and reiterated in Sec.~ \ref{sec:spcpol}, the displacement field  inherits the multivalued nature of $P$. This means the branch shift of $3 \Delta_q P$ as found by aligning the polarization needs to be accounted for when imposing the $D$ value in the constant $D$ simulation. Specifically, the actual branch-matched $D_\rmm{CNC}$ in our case differs from the theoretical target $D_\rmm{CNC}$ by $4\pi (3 \Delta_q P)$, which is by coincidence equal to the theoretical $D_\rmm{CNC}=12 \pi \Delta_q P $ itself. In other words, applying $D=0$, should restore the CNC state for our setup.

\subsubsection{Capacitance and dielectric response}

\begin{figure}[h]
\begin{center}
\resizebox{.48\textwidth}{!}{\includegraphics{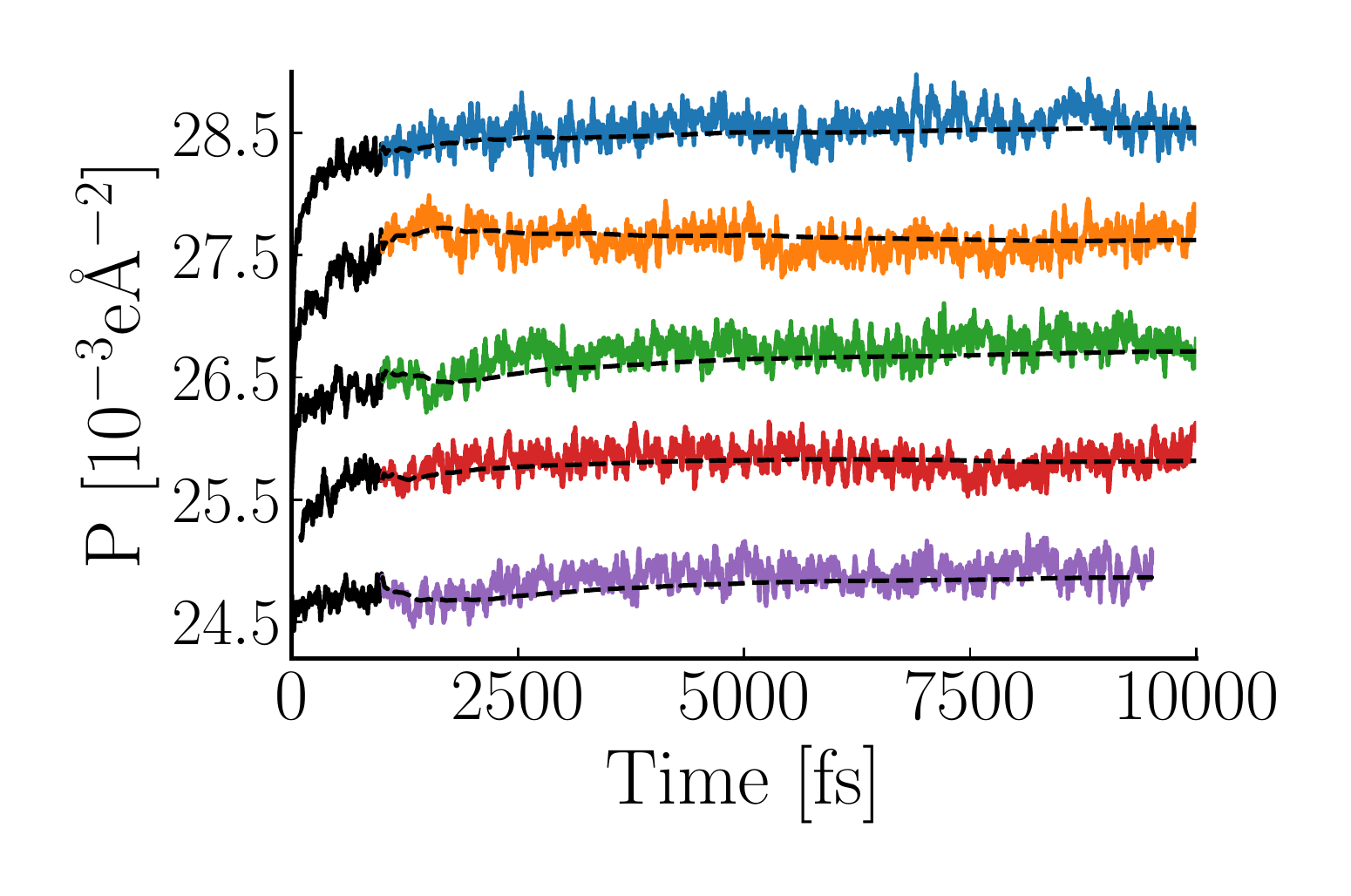}}
\end{center}
\vspace{-25pt}
\caption{Polarization time-series data from the DFTMD trajectories. The first 1~ps was discarded (black). The cumulative average is shown as the black, dashed line. The red (second to last) series had its very early fs behaviour removed as it was very erratic, likely due to the SCF converging on a higher energy state. The blue (top) series crosses the Berry phase boundary during the trajectory and has been unwrapped.} \label{fig:pvst}
\end{figure}

With above considerations in minds, DFTMD at $D=0$ was propagated for $\sim$10~ps (Fig.~\ref{fig:pvst}), with the first 1~ps discarded as an additional equilibration time. The remaining 9~ps were used to calculate $\langle P \rangle$. The polarization was found to be $24.86\times10^{-3}\rmm{e\AA}^{-2}$. Using Eq.~\ref{eq:capacitance} and adjusting the branch shift of $3 \Delta_q P$, one gets a capacitance of 26.38~$\upmu$Fcm$^{-2}$. This is to be compared with the 8.66 $\upmu$Fcm$^{-2}$ from FFMD with the polarization of $26.26\times10^{-3}\rmm{e\AA}^{-2}$.

A factor of three difference between capacitance values may seem counter intuitive because of only a relatively small $\sim$10\% difference in the polarizations. However, this is indeed the case and due to the form of Eq.~\ref{eq:capacitance} in which $\delta C_\text{H}/C_\text{H}$ is unfavourably scaled up by $C_\text{H}$. One might therefore be concerned about the convergence of $C_\rmm{H}$ calculated from DFTMD. To estimate this error, we randomly took 10 uncorrelated windows of 9~ps from FFMD and calculated a standard deviation of $0.044\times10^{-3}\rmm{e\AA}^{-2}$. This translates to 2\% error in the calculated $C_\text{H}$ from FFMD and an estimation of 6\% error in the calculated $C_\text{H}$ from DFTMD.

\begin{figure}[h]
\begin{center}
\resizebox{.48\textwidth}{!}{\includegraphics{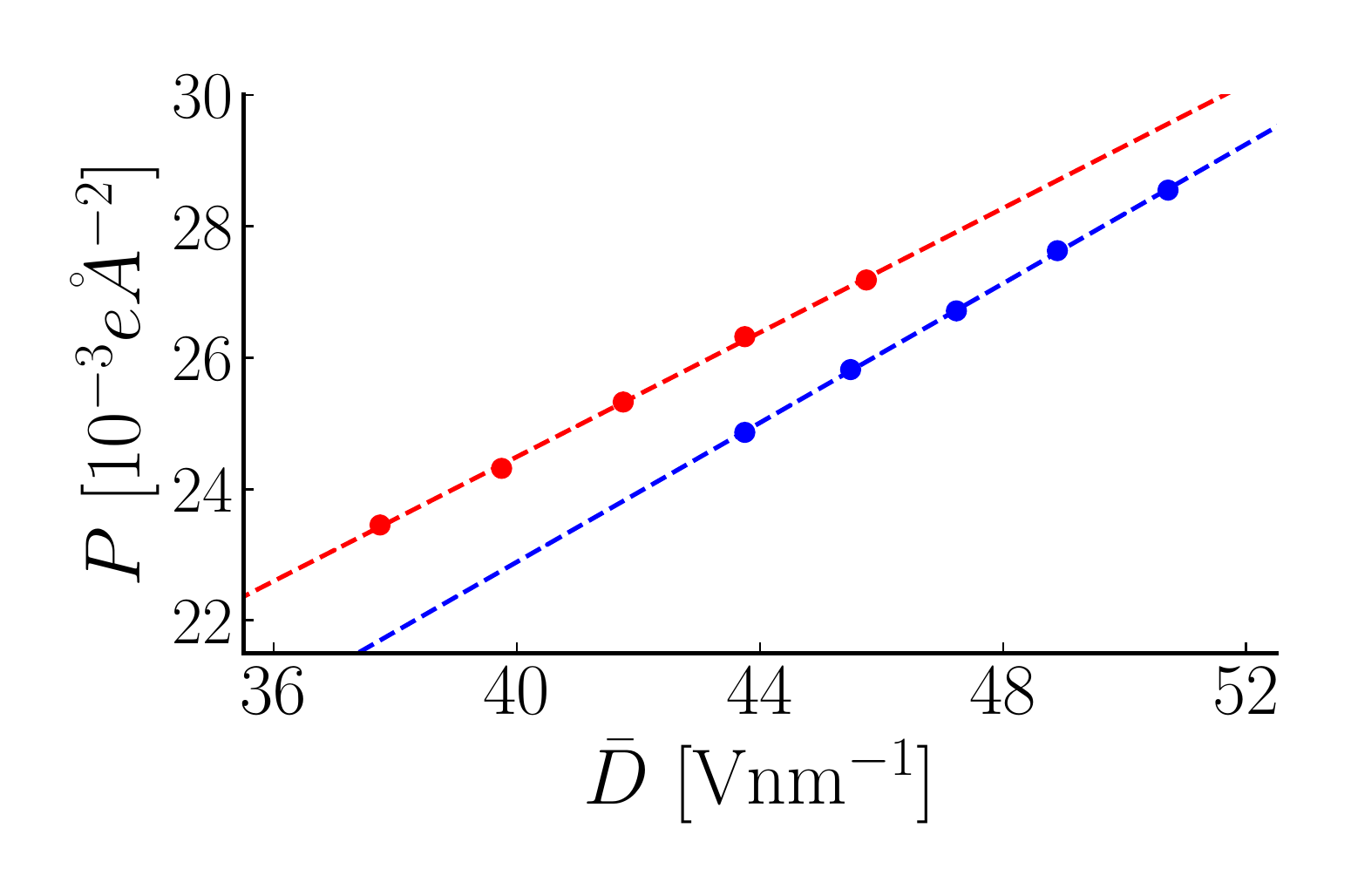}}
\end{center}
\caption{Constant displacement field simulations. CNC is at 43.75~$\rmm{Vnm}^{-1}$. The red (top) series is FFMD, the blue (bottom) series is DFTMD. The gradients are 0.47325 and 0.53010 respectively.} \label{fig:Pcomparison}
\end{figure}

In order to compute ${\epsilon}_\perp$ which serves as the overall dielectric constant of the composite system, we also carried out four additional constant $D$ DFTMD simulations at different $D$ values between zero and the Zener breakdown voltage (Fig.~\ref{fig:pvst}). These data are compared with those obtained from FFMD in Fig.~\ref{fig:Pcomparison}. From Eq.~\ref{eq:Dstate}, the gradient gave ${\epsilon}_\perp$, as 7.6 and 24.5 for FF and DFT systems respectively. The ratio is roughly  the same ratio as for the $C_\text{H}$ estimates, which is just a coincidence.  According to Eq.~\ref{eq:bareps}, ${\epsilon}_\perp$ is determined by the leading term $C_\text{d}$ since the ionic solid and EDLs can be viewed as capacitors connected in series and $n$ is always larger than 2. This is also the reason why the finite size effect which plagues the computation of $C_\text{H}$ is so serious. The charge planes in solid NaCl in FFMD simulation are separated by vacuum ($\epsilon_d = 1$). However DFTMD simulation includes electronic polarization. The optical dielectric constant of NaCl solid is close to 3, as estimated from the refractive index. Thus, it is the electronic polarization which causes the factor 3 difference in ${\epsilon}_\perp$. 

\begin{figure}[h]
\begin{center}
\resizebox{.48\textwidth}{!}{\includegraphics{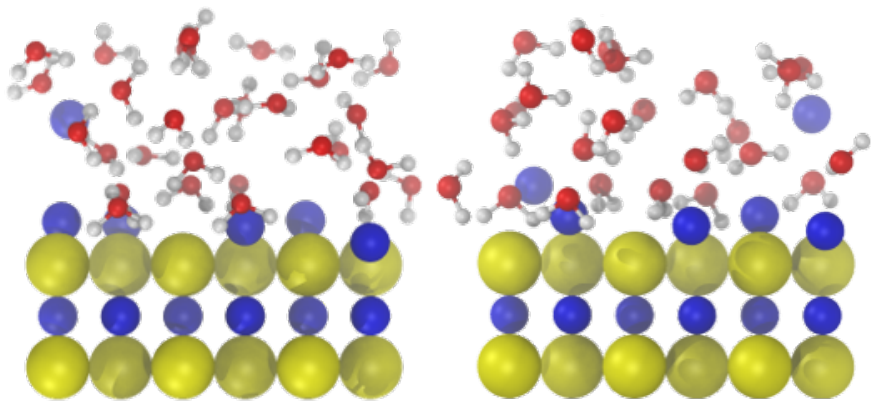}}
\end{center}
\caption{MD snapshots from FFMD (left) and DFTMD (right) simulations of NaCl(111)/NaCl solution system at $D_\text{CNC}$. The FFMD snapshot is the starting geometry for the DFTMD. The snapshot on the right can be seen to have a greater sodium-surface separation, see figure~\ref{fig:comparedensitygraph} for quantitative detail.} \label{fig:comparedensity}
\end{figure}

\begin{figure}[h]
\begin{center}
\resizebox{.48\textwidth}{!}{\includegraphics{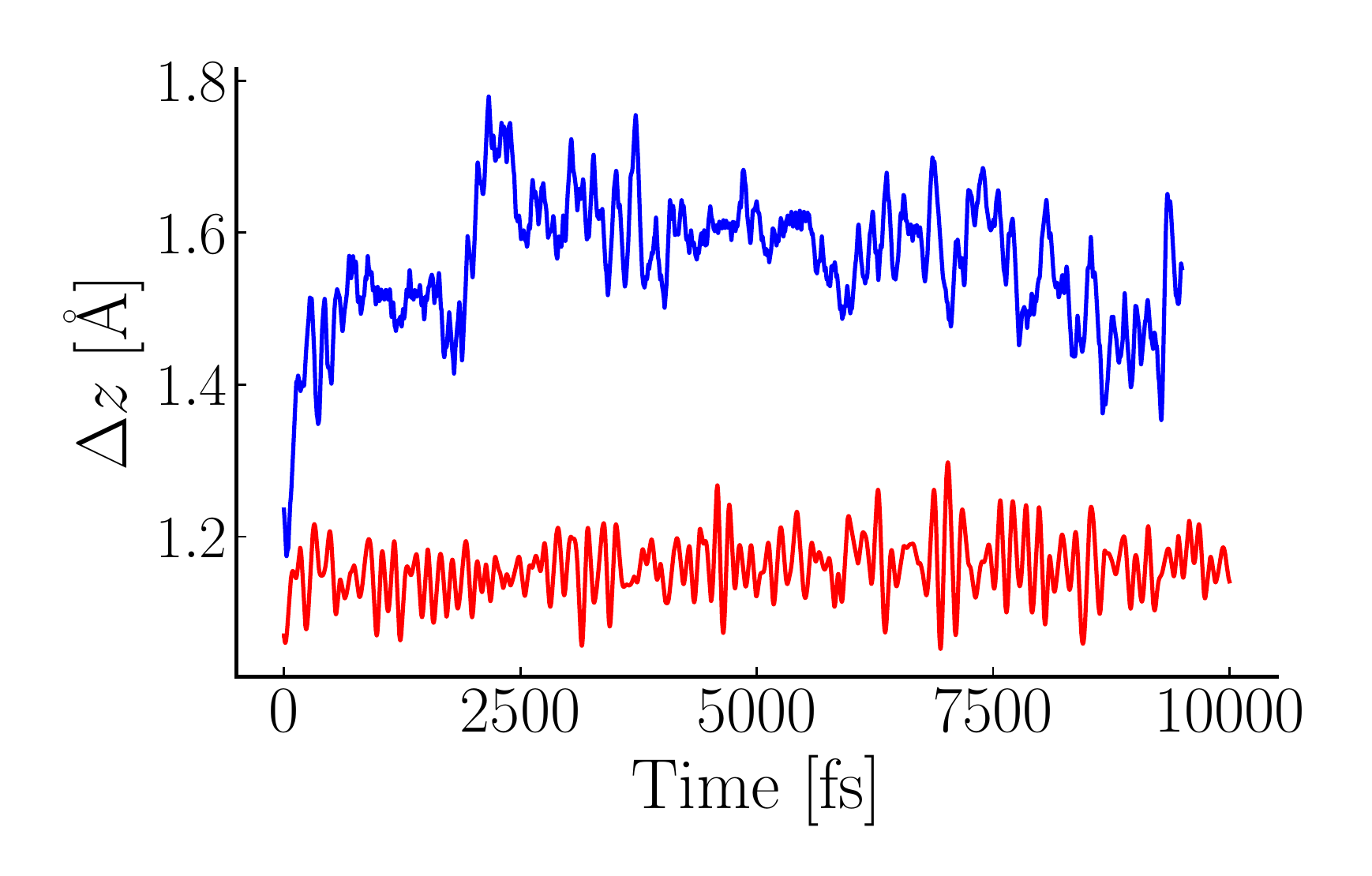}}
\resizebox{.48\textwidth}{!}{\includegraphics{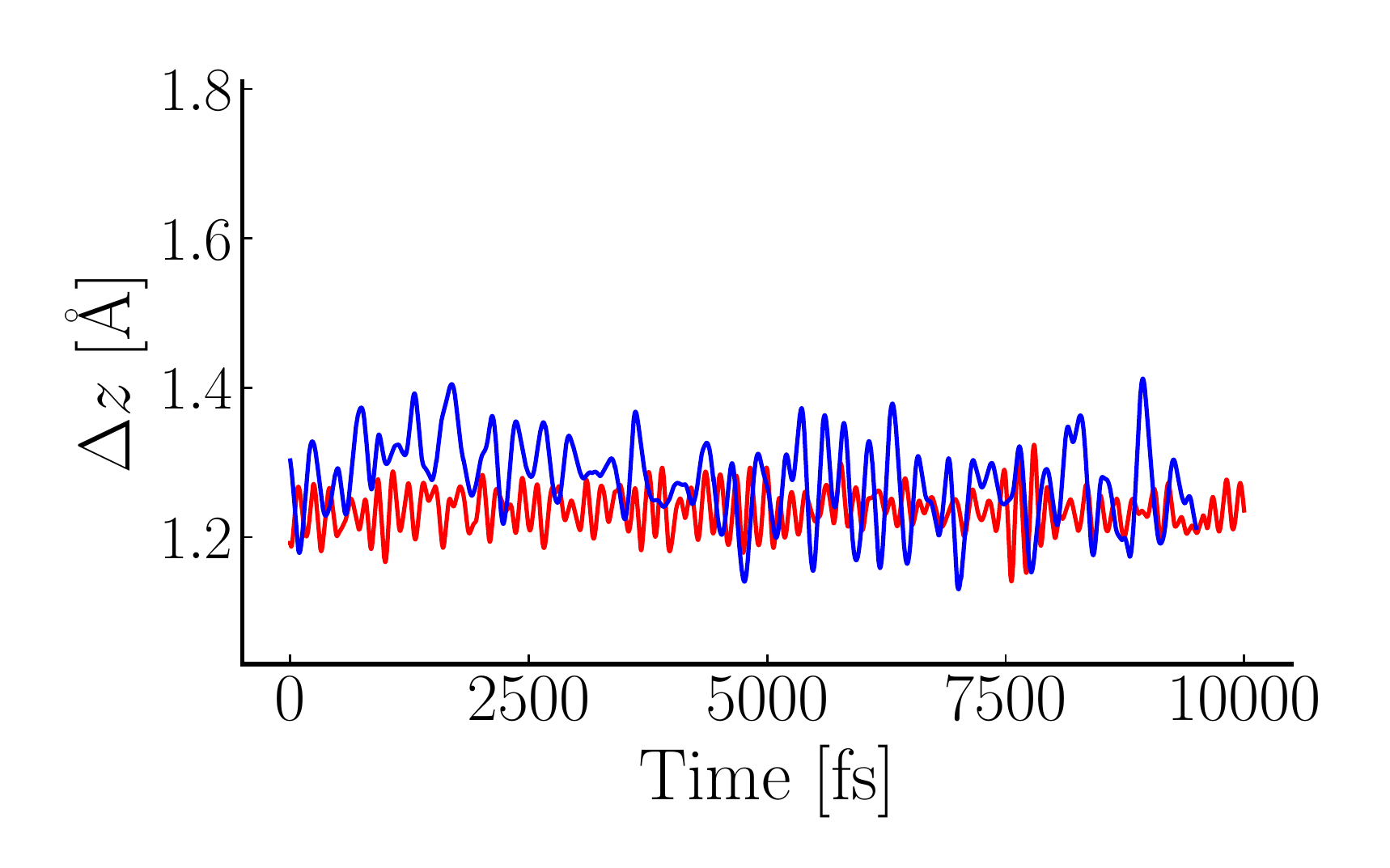}}
\end{center}
\vspace{-20pt}
\caption{Constant displacement field simulations. Top: aqueous sodium, Bottom: aqueous chloride. The red (bottom) series are FFMD, the blue (top) series are DFTMD. The $z$-positions of the 5 inner-shell ions have been averaged and are displayed relative to the fixed surface plane.  Both  FFMD and DFTMD simulations have the same starting geometry other than a short geometry optimization for initiating DFTMD.} \label{fig:comparedensitygraph}
\end{figure}

In paper I the difference in capacitance between the polar NaCl(111) and non-polar NaCl(100) was rationalized in terms of the double layer structure. Here again we see a plane of inner-sphere counter-ions on both sodium and chlorine sides of NaCl(111) surface as the main component of the EDLs (Fig.~\ref{fig:comparedensity}). Distances between the inner-sphere counter-ions and the polar surfaces are plotted in Fig.~\ref{fig:comparedensitygraph}. As one can see, the distance between Cl$^-$(solution) and Na$^+$ (surface) as well the distance between Na$^+$ (solution) and Cl$^-$ (surface) are actually smaller than the layer-wise distance in the ionic crystal in [111] direction which is 1.63\AA. For the case of FFMD simulations, the distances between the inner-sphere counter-ions and NaCl(111) surfaces are about 1.2~\AA. Such a distance in vacuum will lead to a theoretical capacitance of 7.4$~\upmu$Fcm$^{-2}$ which is rather close to the one calculated from the simulation (8.7 $\upmu$Fcm$^{-2}$).  For the case of DFTMD simulation, the distance between Cl$^-$(solution) and Na$^+$ (surface) is quite similar to that from FFMD simulations but the distance between Na$^+$ (solution) is about 50\% larger than that from FFMD simulations. Therefore, the capacitance of polar NaCl(111) calculated from DFTMD simulations should be smaller when compared to that from FFMD if it solely depended on the ionic structure. Instead,  the opposite was found, and $C_\text{H}$ from DFTMD vs. FFMD is 26.4~$\upmu$Fcm$^{-2}$ vs.~8.7~$\upmu$Fcm$^{-2}$. Although there is indeed a contribution to $C_\text{H}$ from outer-sphere counter-ions (Fig.~\ref{fig:comparedensity}), the inner-sphere contribution will be the dominant part in direct analogy to the dead-layer effect of nanoconfined water~\cite{Zhang:2018bn}.  This suggests that it is the electronic polarization present in the DFTMD simulations which determines the final value of $C_\text{H}$ at the NaCl(111)/NaCl solution interface. 

\section{Conclusions} \label{sec:conclu}

Stabilization mechanisms for Type III polar surfaces depend on the physical conditions which the surfaces are exposed to. In case of solid-vacuum and solid-solid interfaces, non-stoichiometric reconstruction and electronic reconstruction are the options. When in contact with electrolytes, such surfaces can be stabilized by supplying the compensating charge in the form of counter-ions from solution, preserving the composition of the solid surface. This is the stabilization mechanism we have investigated here using NaCl(111)/aqueous NaCl solution as a prototype system.

The challenges to study such polar ionic solid/electrolyte systems are two-fold: the inevitable finite size errors of atomistic models and the time-scale needed to convergence the calculation, particularly in the case of DFTMD simulations. We overcame the first challenge in the previous work by imposing a compensating electric field to locate the CNC state and validated the method with FFMD simulations~\cite{Sayer2017}. 

In this work, we expanded our study to DFTMD simulations of the same system and tackled the second challenge with constant electric displacement $D$ simulations. The theoretical formula of $D_\text{CNC}$ which involves only structural parameters (Eq.~\ref{eq:ourD} and associated equations) was first validated against FFMD simulations and then transferred to DFTMD simulations. Despite the fact that the estimator of $C_\text{H}$ of polar surfaces at $D_\text{CNC}$ suffers from unfavourable error propagation, it is feasible to obtain results of a reasonable accuracy within the commonly accessible time-scale of DFTMD (tens of picoseconds).

Comparing results of the Helmholtz capacitance $C_\text{H}$ between FFMD and DFTMD simulations for the same supercell of NaCl(111)/NaCl solution system, it is found that $C_\text{H}$ is dominated by inner-sphere counter-ion contributions with the electronic polarization of ionic solids determining the resulting value. This suggests that DFTMD is indispensable in modelling the charge compensation phenomena and EDLs at polar ionic solid/electrolyte interfaces. This needs to be backed up by a detailed structural analysis which has not been attempted in the present calculation focused on methodology and dielectric response.  Because of its relevance to precipitation and nucleation, this work further hints that the electronic polarization needs to be taken into account when investigating the thermodynamics and kinetics of these processes. A related field where our finite field methods might be of use is that of nano-electrochemistry and nano-ionics\cite{Valov16,Aono16,Kalinin:2018natphys}.

\begin{acknowledgments}
The authors thank J\"urg Hutter (University of Zurich) for getting the CP2K code ready for this application. TS is supported by a departmental studentship (No.~RG84040) sponsored by the Engineering and Sciences Research Council (EPSRC) of the United Kingdom. Computational resources were provided by the UK Car-Parrinello (UKCP) consortium funded by EPSRC. This project was the subject of a HPC-Europa3 grant to TS and we thank the local support from KTH-PDC (Sweden). CZ gratefully acknowledges Uppsala University for the support of a start-up grant and \AA forsk foundation for a research grant (No.~18-460).
\end{acknowledgments}

\bibliography{PolarSurfaces}

\begin{thebibliography}{40}%
\makeatletter
\providecommand \@ifxundefined [1]{%
 \@ifx{#1\undefined}
}%
\providecommand \@ifnum [1]{%
 \ifnum #1\expandafter \@firstoftwo
 \else \expandafter \@secondoftwo
 \fi
}%
\providecommand \@ifx [1]{%
 \ifx #1\expandafter \@firstoftwo
 \else \expandafter \@secondoftwo
 \fi
}%
\providecommand \natexlab [1]{#1}%
\providecommand \enquote  [1]{``#1''}%
\providecommand \bibnamefont  [1]{#1}%
\providecommand \bibfnamefont [1]{#1}%
\providecommand \citenamefont [1]{#1}%
\providecommand \href@noop [0]{\@secondoftwo}%
\providecommand \href [0]{\begingroup \@sanitize@url \@href}%
\providecommand \@href[1]{\@@startlink{#1}\@@href}%
\providecommand \@@href[1]{\endgroup#1\@@endlink}%
\providecommand \@sanitize@url [0]{\catcode `\\12\catcode `\$12\catcode
  `\&12\catcode `\#12\catcode `\^12\catcode `\_12\catcode `\%12\relax}%
\providecommand \@@startlink[1]{}%
\providecommand \@@endlink[0]{}%
\providecommand \url  [0]{\begingroup\@sanitize@url \@url }%
\providecommand \@url [1]{\endgroup\@href {#1}{\urlprefix }}%
\providecommand \urlprefix  [0]{URL }%
\providecommand \Eprint [0]{\href }%
\providecommand \doibase [0]{http://dx.doi.org/}%
\providecommand \selectlanguage [0]{\@gobble}%
\providecommand \bibinfo  [0]{\@secondoftwo}%
\providecommand \bibfield  [0]{\@secondoftwo}%
\providecommand \translation [1]{[#1]}%
\providecommand \BibitemOpen [0]{}%
\providecommand \bibitemStop [0]{}%
\providecommand \bibitemNoStop [0]{.\EOS\space}%
\providecommand \EOS [0]{\spacefactor3000\relax}%
\providecommand \BibitemShut  [1]{\csname bibitem#1\endcsname}%
\let\auto@bib@innerbib\@empty
\bibitem [{\citenamefont {Tasker}(1979)}]{Tasker1979}%
  \BibitemOpen
  \bibfield  {author} {\bibinfo {author} {\bibfnamefont {P.~W.}\ \bibnamefont
  {Tasker}},\ }\href {\doibase 10.1088/0022-3719/12/22/036} {\bibfield
  {journal} {\bibinfo  {journal} {J. Phys. C Solid State Phys.}\ }\textbf
  {\bibinfo {volume} {12}},\ \bibinfo {pages} {4977} (\bibinfo {year}
  {1979})}\BibitemShut {NoStop}%
\bibitem [{\citenamefont {Goniakowski}, \citenamefont {Finocchi},\ and\
  \citenamefont {Noguera}(2008)}]{Goniakowski2008}%
  \BibitemOpen
  \bibfield  {author} {\bibinfo {author} {\bibfnamefont {J.}~\bibnamefont
  {Goniakowski}}, \bibinfo {author} {\bibfnamefont {F.}~\bibnamefont
  {Finocchi}}, \ and\ \bibinfo {author} {\bibfnamefont {C.}~\bibnamefont
  {Noguera}},\ }\href {\doibase 10.1088/0034-4885/71/1/016501} {\bibfield
  {journal} {\bibinfo  {journal} {Reports Prog. Phys.}\ }\textbf {\bibinfo
  {volume} {71}},\ \bibinfo {pages} {016501} (\bibinfo {year}
  {2008})}\BibitemShut {NoStop}%
\bibitem [{\citenamefont {Sayer}, \citenamefont {Zhang},\ and\ \citenamefont
  {Sprik}(2017)}]{Sayer2017}%
  \BibitemOpen
  \bibfield  {author} {\bibinfo {author} {\bibfnamefont {T.}~\bibnamefont
  {Sayer}}, \bibinfo {author} {\bibfnamefont {C.}~\bibnamefont {Zhang}}, \ and\
  \bibinfo {author} {\bibfnamefont {M.}~\bibnamefont {Sprik}},\ }\href@noop {}
  {\bibfield  {journal} {\bibinfo  {journal} {J. Chem. Phys.}\ }\textbf
  {\bibinfo {volume} {147}},\ \bibinfo {pages} {104702} (\bibinfo {year}
  {2017})}\BibitemShut {NoStop}%
\bibitem [{\citenamefont {Noguera}\ and\ \citenamefont
  {Goniakowski}(2013)}]{Noguera2013}%
  \BibitemOpen
  \bibfield  {author} {\bibinfo {author} {\bibfnamefont {C.}~\bibnamefont
  {Noguera}}\ and\ \bibinfo {author} {\bibfnamefont {J.}~\bibnamefont
  {Goniakowski}},\ }\href@noop {} {\bibfield  {journal} {\bibinfo  {journal}
  {Chem. Rev.}\ }\textbf {\bibinfo {volume} {113}},\ \bibinfo {pages} {4073}
  (\bibinfo {year} {2013})}\BibitemShut {NoStop}%
\bibitem [{\citenamefont {Stengel}, \citenamefont {Spaldin},\ and\
  \citenamefont {Vanderbilt}(2009)}]{Stengel2009a}%
  \BibitemOpen
  \bibfield  {author} {\bibinfo {author} {\bibfnamefont {M.}~\bibnamefont
  {Stengel}}, \bibinfo {author} {\bibfnamefont {N.~A.}\ \bibnamefont
  {Spaldin}}, \ and\ \bibinfo {author} {\bibfnamefont {D.}~\bibnamefont
  {Vanderbilt}},\ }\href@noop {} {\bibfield  {journal} {\bibinfo  {journal}
  {Nature Phys.}\ }\textbf {\bibinfo {volume} {5}},\ \bibinfo {pages} {304}
  (\bibinfo {year} {2009})}\BibitemShut {NoStop}%
\bibitem [{\citenamefont {Stengel}, \citenamefont {Vanderbilt},\ and\
  \citenamefont {Spaldin}(2009)}]{Stengel2009b}%
  \BibitemOpen
  \bibfield  {author} {\bibinfo {author} {\bibfnamefont {M.}~\bibnamefont
  {Stengel}}, \bibinfo {author} {\bibfnamefont {D.}~\bibnamefont {Vanderbilt}},
  \ and\ \bibinfo {author} {\bibfnamefont {N.~A.}\ \bibnamefont {Spaldin}},\
  }\href@noop {} {\bibfield  {journal} {\bibinfo  {journal} {Phys.~Rev.~B}\
  }\textbf {\bibinfo {volume} {80}},\ \bibinfo {pages} {224110} (\bibinfo
  {year} {2009})}\BibitemShut {NoStop}%
\bibitem [{\citenamefont {Sprik}(2018)}]{Sprik2018}%
  \BibitemOpen
  \bibfield  {author} {\bibinfo {author} {\bibfnamefont {M.}~\bibnamefont
  {Sprik}},\ }\href@noop {} {\enquote {\bibinfo {title} {{Finite Maxwell Field
  and Electric Displacement Hamiltonians Derived from a Current Dependent
  Lagrangian}},}\ } (\bibinfo {year} {2018}),\ \bibinfo {note}
  {https://doi.org/10.1080/00268976.2018.1431406}\BibitemShut {NoStop}%
\bibitem [{\citenamefont {Zhang}\ and\ \citenamefont
  {Sprik}(2016{\natexlab{a}})}]{Zhang2016b}%
  \BibitemOpen
  \bibfield  {author} {\bibinfo {author} {\bibfnamefont {C.}~\bibnamefont
  {Zhang}}\ and\ \bibinfo {author} {\bibfnamefont {M.}~\bibnamefont {Sprik}},\
  }\href@noop {} {\bibfield  {journal} {\bibinfo  {journal} {Phys. Rev. B.}\
  }\textbf {\bibinfo {volume} {94}},\ \bibinfo {pages} {245309} (\bibinfo
  {year} {2016}{\natexlab{a}})}\BibitemShut {NoStop}%
\bibitem [{\citenamefont {King-Smith}\ and\ \citenamefont
  {Vanderbilt}(1993)}]{King-Smith:1993prb}%
  \BibitemOpen
  \bibfield  {author} {\bibinfo {author} {\bibfnamefont {R.~D.}\ \bibnamefont
  {King-Smith}}\ and\ \bibinfo {author} {\bibfnamefont {D.}~\bibnamefont
  {Vanderbilt}},\ }\href@noop {} {\bibfield  {journal} {\bibinfo  {journal}
  {Phys.~Rev.~B}\ }\textbf {\bibinfo {volume} {47}},\ \bibinfo {pages} {1651}
  (\bibinfo {year} {1993})}\BibitemShut {NoStop}%
\bibitem [{\citenamefont {Resta}(1994)}]{Resta:1994rmp}%
  \BibitemOpen
  \bibfield  {author} {\bibinfo {author} {\bibfnamefont {R.}~\bibnamefont
  {Resta}},\ }\href@noop {} {\bibfield  {journal} {\bibinfo  {journal}
  {Rev.~Mod.~Phys}\ }\textbf {\bibinfo {volume} {66}},\ \bibinfo {pages} {899}
  (\bibinfo {year} {1994})}\BibitemShut {NoStop}%
\bibitem [{\citenamefont {Resta}\ and\ \citenamefont
  {Vanderbilt}(2007)}]{Resta:2007ch}%
  \BibitemOpen
  \bibfield  {author} {\bibinfo {author} {\bibfnamefont {R.}~\bibnamefont
  {Resta}}\ and\ \bibinfo {author} {\bibfnamefont {D.}~\bibnamefont
  {Vanderbilt}},\ }in\ \href@noop {} {\emph {\bibinfo {booktitle} {{Topics in
  Applied Physics Volume 105: Physics of Ferroelectrics: a Modern
  Perspective}}}},\ \bibinfo {editor} {edited by\ \bibinfo {editor}
  {\bibfnamefont {K.~M.}\ \bibnamefont {Rabe}}, \bibinfo {editor}
  {\bibfnamefont {C.~H.}\ \bibnamefont {Ahn}}, \ and\ \bibinfo {editor}
  {\bibfnamefont {J.-M.}\ \bibnamefont {Triscone}}}\ (\bibinfo  {publisher}
  {Springer-Verlag},\ \bibinfo {year} {2007})\ pp.\ \bibinfo {pages}
  {31--67}\BibitemShut {NoStop}%
\bibitem [{\citenamefont {Zhang}\ and\ \citenamefont
  {Sprik}(2016{\natexlab{b}})}]{Zhang2016a}%
  \BibitemOpen
  \bibfield  {author} {\bibinfo {author} {\bibfnamefont {C.}~\bibnamefont
  {Zhang}}\ and\ \bibinfo {author} {\bibfnamefont {M.}~\bibnamefont {Sprik}},\
  }\href@noop {} {\bibfield  {journal} {\bibinfo  {journal} {Phys. Rev. B.}\
  }\textbf {\bibinfo {volume} {93}},\ \bibinfo {pages} {144201} (\bibinfo
  {year} {2016}{\natexlab{b}})}\BibitemShut {NoStop}%
\bibitem [{\citenamefont {Caillol}(1994)}]{Caillol1994}%
  \BibitemOpen
  \bibfield  {author} {\bibinfo {author} {\bibfnamefont {J.-M.}\ \bibnamefont
  {Caillol}},\ }\href@noop {} {\bibfield  {journal} {\bibinfo  {journal} {J.
  Chem. Phys.}\ }\textbf {\bibinfo {volume} {101}},\ \bibinfo {pages} {6080}
  (\bibinfo {year} {1994})}\BibitemShut {NoStop}%
\bibitem [{\citenamefont {De~Leeuw}, \citenamefont {Perram},\ and\
  \citenamefont {Smith}(1980)}]{DeLeeuw1980a}%
  \BibitemOpen
  \bibfield  {author} {\bibinfo {author} {\bibfnamefont {S.~W.}\ \bibnamefont
  {De~Leeuw}}, \bibinfo {author} {\bibfnamefont {J.~W.}\ \bibnamefont
  {Perram}}, \ and\ \bibinfo {author} {\bibfnamefont {E.~R.}\ \bibnamefont
  {Smith}},\ }\href@noop {} {\bibfield  {journal} {\bibinfo  {journal} {Proc.
  R. Soc. A}\ }\textbf {\bibinfo {volume} {373}},\ \bibinfo {pages} {27}
  (\bibinfo {year} {1980})}\BibitemShut {NoStop}%
\bibitem [{\citenamefont {Caillol}, \citenamefont {Levesque},\ and\
  \citenamefont {Weis}(1989{\natexlab{a}})}]{Caillol1989a}%
  \BibitemOpen
  \bibfield  {author} {\bibinfo {author} {\bibfnamefont {J.-M.}\ \bibnamefont
  {Caillol}}, \bibinfo {author} {\bibfnamefont {D.}~\bibnamefont {Levesque}}, \
  and\ \bibinfo {author} {\bibfnamefont {J.~J.}\ \bibnamefont {Weis}},\
  }\href@noop {} {\bibfield  {journal} {\bibinfo  {journal} {J. Chem. Phys.}\
  }\textbf {\bibinfo {volume} {91}},\ \bibinfo {pages} {5544} (\bibinfo {year}
  {1989}{\natexlab{a}})}\BibitemShut {NoStop}%
\bibitem [{\citenamefont {Caillol}, \citenamefont {Levesque},\ and\
  \citenamefont {Weis}(1989{\natexlab{b}})}]{Caillol1989b}%
  \BibitemOpen
  \bibfield  {author} {\bibinfo {author} {\bibfnamefont {J.-M.}\ \bibnamefont
  {Caillol}}, \bibinfo {author} {\bibfnamefont {D.}~\bibnamefont {Levesque}}, \
  and\ \bibinfo {author} {\bibfnamefont {J.~J.}\ \bibnamefont {Weis}},\
  }\href@noop {} {\bibfield  {journal} {\bibinfo  {journal} {J. Chem. Phys.}\
  }\textbf {\bibinfo {volume} {91}},\ \bibinfo {pages} {5555} (\bibinfo {year}
  {1989}{\natexlab{b}})}\BibitemShut {NoStop}%
\bibitem [{\citenamefont {Martin}(1974)}]{Martin1974}%
  \BibitemOpen
  \bibfield  {author} {\bibinfo {author} {\bibfnamefont {R.~M.}\ \bibnamefont
  {Martin}},\ }\href@noop {} {\bibfield  {journal} {\bibinfo  {journal}
  {Phys.~Rev.~B}\ }\textbf {\bibinfo {volume} {9}},\ \bibinfo {pages} {1998}
  (\bibinfo {year} {1974})}\BibitemShut {NoStop}%
\bibitem [{Note1()}]{Note1}%
  \BibitemOpen
  \bibinfo {note} {This is Eq. 6 of Paper I, which contained a
  typo.}\BibitemShut {Stop}%
\bibitem [{\citenamefont {Resta}(1998)}]{Resta:1998prl}%
  \BibitemOpen
  \bibfield  {author} {\bibinfo {author} {\bibfnamefont {R.}~\bibnamefont
  {Resta}},\ }\href@noop {} {\bibfield  {journal} {\bibinfo  {journal}
  {Phys.~Rev.~Lett.}\ }\textbf {\bibinfo {volume} {80}},\ \bibinfo {pages}
  {1800} (\bibinfo {year} {1998})}\BibitemShut {NoStop}%
\bibitem [{\citenamefont {Resta}(2000)}]{Resta:2000jpcm}%
  \BibitemOpen
  \bibfield  {author} {\bibinfo {author} {\bibfnamefont {R.}~\bibnamefont
  {Resta}},\ }\href@noop {} {\bibfield  {journal} {\bibinfo  {journal}
  {J.~Phys.: Condens.~Matter}\ }\textbf {\bibinfo {volume} {12}},\ \bibinfo
  {pages} {R107} (\bibinfo {year} {2000})}\BibitemShut {NoStop}%
\bibitem [{\citenamefont {Hutter}\ \emph {et~al.}(2014)\citenamefont {Hutter},
  \citenamefont {Iannuzzi}, \citenamefont {Schiffmann},\ and\ \citenamefont
  {VandeVondele}}]{Hutter:2013iea}%
  \BibitemOpen
  \bibfield  {author} {\bibinfo {author} {\bibfnamefont {J.}~\bibnamefont
  {Hutter}}, \bibinfo {author} {\bibfnamefont {M.}~\bibnamefont {Iannuzzi}},
  \bibinfo {author} {\bibfnamefont {F.}~\bibnamefont {Schiffmann}}, \ and\
  \bibinfo {author} {\bibfnamefont {J.}~\bibnamefont {VandeVondele}},\
  }\href@noop {} {\bibfield  {journal} {\bibinfo  {journal} {WIREs Comput. Mol.
  Sci.}\ }\textbf {\bibinfo {volume} {4}},\ \bibinfo {pages} {15} (\bibinfo
  {year} {2014})}\BibitemShut {NoStop}%
\bibitem [{\citenamefont {Marzari}\ and\ \citenamefont
  {Vanderbilt}(1997)}]{Marzari:1997wa}%
  \BibitemOpen
  \bibfield  {author} {\bibinfo {author} {\bibfnamefont {N.}~\bibnamefont
  {Marzari}}\ and\ \bibinfo {author} {\bibfnamefont {D.}~\bibnamefont
  {Vanderbilt}},\ }\href@noop {} {\bibfield  {journal} {\bibinfo  {journal}
  {Phys.~Rev.~B}\ }\textbf {\bibinfo {volume} {56}},\ \bibinfo {pages} {12847}
  (\bibinfo {year} {1997})}\BibitemShut {NoStop}%
\bibitem [{\citenamefont {Silvestrelli}\ \emph {et~al.}(1998)\citenamefont
  {Silvestrelli}, \citenamefont {Marzari}, \citenamefont {Vanderbilt},\ and\
  \citenamefont {Parrinello}}]{Silvestrelli:1998ssc}%
  \BibitemOpen
  \bibfield  {author} {\bibinfo {author} {\bibfnamefont {P.~L.}\ \bibnamefont
  {Silvestrelli}}, \bibinfo {author} {\bibfnamefont {N.}~\bibnamefont
  {Marzari}}, \bibinfo {author} {\bibfnamefont {D.}~\bibnamefont {Vanderbilt}},
  \ and\ \bibinfo {author} {\bibfnamefont {M.}~\bibnamefont {Parrinello}},\
  }\href@noop {} {\bibfield  {journal} {\bibinfo  {journal} {Solid State
  Commun.}\ }\textbf {\bibinfo {volume} {107}},\ \bibinfo {pages} {7} (\bibinfo
  {year} {1998})}\BibitemShut {NoStop}%
\bibitem [{\citenamefont {Marzari}\ \emph {et~al.}(2012)\citenamefont
  {Marzari}, \citenamefont {Mostofi}, \citenamefont {Yates}, \citenamefont
  {Souza},\ and\ \citenamefont {Vanderbilt}}]{Marzari:2012eu}%
  \BibitemOpen
  \bibfield  {author} {\bibinfo {author} {\bibfnamefont {N.}~\bibnamefont
  {Marzari}}, \bibinfo {author} {\bibfnamefont {A.~A.}\ \bibnamefont
  {Mostofi}}, \bibinfo {author} {\bibfnamefont {J.~R.}\ \bibnamefont {Yates}},
  \bibinfo {author} {\bibfnamefont {.~I.}\ \bibnamefont {Souza}}, \ and\
  \bibinfo {author} {\bibfnamefont {D.}~\bibnamefont {Vanderbilt}},\
  }\href@noop {} {\bibfield  {journal} {\bibinfo  {journal} {Rev.~Mod/~Phys.}\
  }\textbf {\bibinfo {volume} {84}},\ \bibinfo {pages} {1419} (\bibinfo {year}
  {2012})}\BibitemShut {NoStop}%
\bibitem [{\citenamefont {Hess}\ \emph {et~al.}(2008)\citenamefont {Hess},
  \citenamefont {Kutzner}, \citenamefont {van~der Spoel},\ and\ \citenamefont
  {Lindahl}}]{Hess2008}%
  \BibitemOpen
  \bibfield  {author} {\bibinfo {author} {\bibfnamefont {B.}~\bibnamefont
  {Hess}}, \bibinfo {author} {\bibfnamefont {C.}~\bibnamefont {Kutzner}},
  \bibinfo {author} {\bibfnamefont {D.}~\bibnamefont {van~der Spoel}}, \ and\
  \bibinfo {author} {\bibfnamefont {E.}~\bibnamefont {Lindahl}},\ }\href@noop
  {} {\bibfield  {journal} {\bibinfo  {journal} {J. Chem. Theory Comput.}\
  }\textbf {\bibinfo {volume} {4}},\ \bibinfo {pages} {435} (\bibinfo {year}
  {2008})}\BibitemShut {NoStop}%
\bibitem [{\citenamefont {Berendsen}, \citenamefont {Grigera},\ and\
  \citenamefont {Straatsma}(1987)}]{Berendsen:1987uu}%
  \BibitemOpen
  \bibfield  {author} {\bibinfo {author} {\bibfnamefont {H.~J.~C.}\
  \bibnamefont {Berendsen}}, \bibinfo {author} {\bibfnamefont {J.~R.}\
  \bibnamefont {Grigera}}, \ and\ \bibinfo {author} {\bibfnamefont {T.~P.}\
  \bibnamefont {Straatsma}},\ }\href@noop {} {\bibfield  {journal} {\bibinfo
  {journal} {J. Phys. Chem.}\ }\textbf {\bibinfo {volume} {91}},\ \bibinfo
  {pages} {6269} (\bibinfo {year} {1987})}\BibitemShut {NoStop}%
\bibitem [{\citenamefont {Joung}\ and\ \citenamefont
  {T~E~Cheatham}(2008)}]{Joung2008}%
  \BibitemOpen
  \bibfield  {author} {\bibinfo {author} {\bibfnamefont {I.~S.}\ \bibnamefont
  {Joung}}\ and\ \bibinfo {author} {\bibfnamefont {I.}~\bibnamefont
  {T~E~Cheatham}},\ }\href@noop {} {\bibfield  {journal} {\bibinfo  {journal}
  {J. Phys. Chem. B}\ }\textbf {\bibinfo {volume} {112}},\ \bibinfo {pages}
  {9020} (\bibinfo {year} {2008})}\BibitemShut {NoStop}%
\bibitem [{\citenamefont {Zhang}\ \emph {et~al.}(2010)\citenamefont {Zhang},
  \citenamefont {Raugei}, \citenamefont {Eisenberg},\ and\ \citenamefont
  {Carloni}}]{Zhang:2010zh}%
  \BibitemOpen
  \bibfield  {author} {\bibinfo {author} {\bibfnamefont {C.}~\bibnamefont
  {Zhang}}, \bibinfo {author} {\bibfnamefont {S.}~\bibnamefont {Raugei}},
  \bibinfo {author} {\bibfnamefont {B.}~\bibnamefont {Eisenberg}}, \ and\
  \bibinfo {author} {\bibfnamefont {P.}~\bibnamefont {Carloni}},\ }\href@noop
  {} {\bibfield  {journal} {\bibinfo  {journal} {J. Chem. Theory Comput.}\
  }\textbf {\bibinfo {volume} {6}},\ \bibinfo {pages} {2167} (\bibinfo {year}
  {2010})}\BibitemShut {NoStop}%
\bibitem [{\citenamefont {Nezbeda}, \citenamefont {Moučka},\ and\
  \citenamefont {Smith}(2016)}]{Nezbeda2016}%
  \BibitemOpen
  \bibfield  {author} {\bibinfo {author} {\bibfnamefont {I.}~\bibnamefont
  {Nezbeda}}, \bibinfo {author} {\bibfnamefont {F.}~\bibnamefont {Moučka}}, \
  and\ \bibinfo {author} {\bibfnamefont {W.~R.}\ \bibnamefont {Smith}},\
  }\href@noop {} {\bibfield  {journal} {\bibinfo  {journal} {Mol. Phys.}\
  }\textbf {\bibinfo {volume} {114}},\ \bibinfo {pages} {1665} (\bibinfo {year}
  {2016})}\BibitemShut {NoStop}%
\bibitem [{\citenamefont {VandeVondele}\ \emph {et~al.}(2005)\citenamefont
  {VandeVondele}, \citenamefont {Krack}, \citenamefont {Mohamed}, \citenamefont
  {Parrinello}, \citenamefont {Chassaing},\ and\ \citenamefont
  {Hutter}}]{VandeVondele:2005ge}%
  \BibitemOpen
  \bibfield  {author} {\bibinfo {author} {\bibfnamefont {J.}~\bibnamefont
  {VandeVondele}}, \bibinfo {author} {\bibfnamefont {M.}~\bibnamefont {Krack}},
  \bibinfo {author} {\bibfnamefont {F.}~\bibnamefont {Mohamed}}, \bibinfo
  {author} {\bibfnamefont {M.}~\bibnamefont {Parrinello}}, \bibinfo {author}
  {\bibfnamefont {T.}~\bibnamefont {Chassaing}}, \ and\ \bibinfo {author}
  {\bibfnamefont {J.}~\bibnamefont {Hutter}},\ }\href@noop {} {\bibfield
  {journal} {\bibinfo  {journal} {Comp. Phys. Comm.}\ }\textbf {\bibinfo
  {volume} {167}},\ \bibinfo {pages} {103} (\bibinfo {year}
  {2005})}\BibitemShut {NoStop}%
\bibitem [{\citenamefont {Goedecker}, \citenamefont {Teter},\ and\
  \citenamefont {Hutter}(1996)}]{Goedecker:1996ve}%
  \BibitemOpen
  \bibfield  {author} {\bibinfo {author} {\bibfnamefont {S.}~\bibnamefont
  {Goedecker}}, \bibinfo {author} {\bibfnamefont {M.}~\bibnamefont {Teter}}, \
  and\ \bibinfo {author} {\bibfnamefont {J.}~\bibnamefont {Hutter}},\
  }\href@noop {} {\bibfield  {journal} {\bibinfo  {journal} {Phys.~Rev.B}\
  }\textbf {\bibinfo {volume} {54}},\ \bibinfo {pages} {1703} (\bibinfo {year}
  {1996})}\BibitemShut {NoStop}%
\bibitem [{\citenamefont {VandeVondele}\ and\ \citenamefont
  {Hutter}(2007)}]{VandeVondele:2007wt}%
  \BibitemOpen
  \bibfield  {author} {\bibinfo {author} {\bibfnamefont {J.}~\bibnamefont
  {VandeVondele}}\ and\ \bibinfo {author} {\bibfnamefont {J.}~\bibnamefont
  {Hutter}},\ }\href@noop {} {\bibfield  {journal} {\bibinfo  {journal} {J.
  Chem. Phys.}\ }\textbf {\bibinfo {volume} {127}},\ \bibinfo {pages} {114105}
  (\bibinfo {year} {2007})}\BibitemShut {NoStop}%
\bibitem [{\citenamefont {Perdew}, \citenamefont {Burke},\ and\ \citenamefont
  {Ernzerhof}(1996)}]{PhysRevLett.77.3865}%
  \BibitemOpen
  \bibfield  {author} {\bibinfo {author} {\bibfnamefont {J.~P.}\ \bibnamefont
  {Perdew}}, \bibinfo {author} {\bibfnamefont {K.}~\bibnamefont {Burke}}, \
  and\ \bibinfo {author} {\bibfnamefont {M.}~\bibnamefont {Ernzerhof}},\
  }\href@noop {} {\bibfield  {journal} {\bibinfo  {journal} {Phys. Rev. Lett.}\
  }\textbf {\bibinfo {volume} {77}},\ \bibinfo {pages} {3865} (\bibinfo {year}
  {1996})}\BibitemShut {NoStop}%
\bibitem [{\citenamefont {Bussi}, \citenamefont {Donadio},\ and\ \citenamefont
  {Parrinello}(2007)}]{Bussi:2008wu}%
  \BibitemOpen
  \bibfield  {author} {\bibinfo {author} {\bibfnamefont {G.}~\bibnamefont
  {Bussi}}, \bibinfo {author} {\bibfnamefont {D.}~\bibnamefont {Donadio}}, \
  and\ \bibinfo {author} {\bibfnamefont {M.}~\bibnamefont {Parrinello}},\
  }\href@noop {} {\bibfield  {journal} {\bibinfo  {journal} {J. Chem. Phys.}\
  }\textbf {\bibinfo {volume} {126}},\ \bibinfo {pages} {014101} (\bibinfo
  {year} {2007})}\BibitemShut {NoStop}%
\bibitem [{\citenamefont {VandeVondele}\ and\ \citenamefont
  {Hutter}(2003)}]{VandeVondele:2003ue}%
  \BibitemOpen
  \bibfield  {author} {\bibinfo {author} {\bibfnamefont {J.}~\bibnamefont
  {VandeVondele}}\ and\ \bibinfo {author} {\bibfnamefont {J.}~\bibnamefont
  {Hutter}},\ }\href@noop {} {\bibfield  {journal} {\bibinfo  {journal} {J.
  Chem. Phys.}\ }\textbf {\bibinfo {volume} {118}},\ \bibinfo {pages} {4365}
  (\bibinfo {year} {2003})}\BibitemShut {NoStop}%
\bibitem [{\citenamefont {Zhang}, \citenamefont {Hutter},\ and\ \citenamefont
  {Sprik}(2016)}]{Zhang2016c}%
  \BibitemOpen
  \bibfield  {author} {\bibinfo {author} {\bibfnamefont {C.}~\bibnamefont
  {Zhang}}, \bibinfo {author} {\bibfnamefont {J.}~\bibnamefont {Hutter}}, \
  and\ \bibinfo {author} {\bibfnamefont {M.}~\bibnamefont {Sprik}},\
  }\href@noop {} {\bibfield  {journal} {\bibinfo  {journal} {J. Phys. Chem.
  Lett.}\ }\textbf {\bibinfo {volume} {7}},\ \bibinfo {pages} {2696} (\bibinfo
  {year} {2016})}\BibitemShut {NoStop}%
\bibitem [{\citenamefont {Zhang}(2018)}]{Zhang:2018bn}%
  \BibitemOpen
  \bibfield  {author} {\bibinfo {author} {\bibfnamefont {C.}~\bibnamefont
  {Zhang}},\ }\href@noop {} {\bibfield  {journal} {\bibinfo  {journal} {J.
  Chem. Phys.}\ }\textbf {\bibinfo {volume} {148}},\ \bibinfo {pages} {156101}
  (\bibinfo {year} {2018})}\BibitemShut {NoStop}%
\bibitem [{\citenamefont {Valov}\ and\ \citenamefont {Lu}(2016)}]{Valov16}%
  \BibitemOpen
  \bibfield  {author} {\bibinfo {author} {\bibfnamefont {I.}~\bibnamefont
  {Valov}}\ and\ \bibinfo {author} {\bibfnamefont {W.~D.}\ \bibnamefont {Lu}},\
  }\href@noop {} {\bibfield  {journal} {\bibinfo  {journal} {Nanoscale}\
  }\textbf {\bibinfo {volume} {8}},\ \bibinfo {pages} {13828} (\bibinfo {year}
  {2016})}\BibitemShut {NoStop}%
\bibitem [{\citenamefont {Terabe}\ \emph {et~al.}(2016)\citenamefont {Terabe},
  \citenamefont {Tsuchiya}, \citenamefont {Yang},\ and\ \citenamefont
  {Aono}}]{Aono16}%
  \BibitemOpen
  \bibfield  {author} {\bibinfo {author} {\bibfnamefont {K.}~\bibnamefont
  {Terabe}}, \bibinfo {author} {\bibfnamefont {T.}~\bibnamefont {Tsuchiya}},
  \bibinfo {author} {\bibfnamefont {R.}~\bibnamefont {Yang}}, \ and\ \bibinfo
  {author} {\bibfnamefont {M.}~\bibnamefont {Aono}},\ }\href@noop {} {\bibfield
   {journal} {\bibinfo  {journal} {Nanoscale}\ }\textbf {\bibinfo {volume}
  {8}},\ \bibinfo {pages} {13873} (\bibinfo {year} {2016})}\BibitemShut
  {NoStop}%
\bibitem [{\citenamefont {Yang}\ \emph {et~al.}(2017)\citenamefont {Yang},
  \citenamefont {Morozovska}, \citenamefont {Kumar}, \citenamefont {Eliseev},
  \citenamefont {Cao}, \citenamefont {Mazet}, \citenamefont {Balke},
  \citenamefont {Jesse}, \citenamefont {Vasudevan}, \citenamefont
  {Dubourdieu},\ and\ \citenamefont {Kalinin}}]{Kalinin:2018natphys}%
  \BibitemOpen
  \bibfield  {author} {\bibinfo {author} {\bibfnamefont {S.~M.}\ \bibnamefont
  {Yang}}, \bibinfo {author} {\bibfnamefont {A.~N.}\ \bibnamefont
  {Morozovska}}, \bibinfo {author} {\bibfnamefont {R.}~\bibnamefont {Kumar}},
  \bibinfo {author} {\bibfnamefont {E.~A.}\ \bibnamefont {Eliseev}}, \bibinfo
  {author} {\bibfnamefont {Y.}~\bibnamefont {Cao}}, \bibinfo {author}
  {\bibfnamefont {L.}~\bibnamefont {Mazet}}, \bibinfo {author} {\bibfnamefont
  {N.}~\bibnamefont {Balke}}, \bibinfo {author} {\bibfnamefont
  {S.}~\bibnamefont {Jesse}}, \bibinfo {author} {\bibfnamefont {R.~K.}\
  \bibnamefont {Vasudevan}}, \bibinfo {author} {\bibfnamefont {C.}~\bibnamefont
  {Dubourdieu}}, \ and\ \bibinfo {author} {\bibfnamefont {S.~V.}\ \bibnamefont
  {Kalinin}},\ }\href@noop {} {\bibfield  {journal} {\bibinfo  {journal}
  {Nat.~Phys.}\ }\textbf {\bibinfo {volume} {13}},\ \bibinfo {pages} {812}
  (\bibinfo {year} {2017})}\BibitemShut {NoStop}%
\end{thebibliography}%

\end{document}